\documentclass[
 reprint,
 superscriptaddress,
 nobibnotes,
 amsmath,amssymb,
 aps,
 floatfix,
]{revtex4-2}

\usepackage{mathtools,xspace}
\usepackage{graphicx}
\usepackage{dcolumn}
\usepackage{bm}
\usepackage{xcolor}
\usepackage{subcaption}
\usepackage{hyperref}
\usepackage{chemformula}
\usepackage{adjustbox}
\usepackage{gensymb}
\usepackage{booktabs}
\usepackage[normalem]{ulem}

\newcommand{\editor}[2]{%
  \expandafter\newcommand\csname #1note\endcsname[1]{%
    \textcolor{#2}{(\textbf{#1:} ##1)}}%
  \expandafter\newcommand\csname #1\endcsname[1]{%
    \textcolor{#2}{##1}}%
  \expandafter\newcommand\csname #1cancel\endcsname[1]{%
    \textcolor{#2}{\sout{##1}}}%
  \expandafter\newcommand\csname #1change\endcsname[2]{%
    \textcolor{#2}{\sout{##1} ##2}}%
  \newenvironment{#1text}{\color{#2}}{\color{black}}
}

\editor{LB}{blue}

\newcommand{\UCB}{
  Department of Materials Science and Engineering,
  University of California Berkeley,
  Berkeley CA, USA
}
\newcommand{\LBL}{
  Materials Sciences Division,
  Lawrence Berkeley National Laboratory,
  Berkeley CA, USA
}
\newcommand{\EPFL}{Theory and Simulation of Materials (THEOS) and National Center for Computational Design and Discovery of Novel Materials (MARVEL), École Polytechnique Fédérale de Lausanne, CH-1015, Lausanne, Switzerland}

\newcommand{\rrscan}{r$^2$SCAN\xspace}

\begin{document}

\title{The Interplay Between Electron Localization, Magnetic Order, and Jahn-Teller Distortion that Dictates \ch{LiMnO2} Phase Stability}

\author{Ronald L. Kam}
\author{Luca Binci}
\affiliation{\UCB}
\affiliation{\LBL}
\author{Aaron D. Kaplan}
\affiliation{\LBL}
\author{Kristin A. Persson}
\affiliation{\UCB}
\affiliation{\LBL}

\author{Nicola Marzari}
\affiliation{\EPFL}

\author{Gerbrand Ceder}
\email{gceder@berkeley.edu}
\affiliation{\UCB}
\affiliation{\LBL}

\date{\today}

\begin{abstract}
The development of manganese (Mn)-rich cathodes for Li-ion batteries promises to alleviate potential supply chain bottlenecks in battery manufacturing. Fundamental challenges in Mn-rich cathodes arise from phenomena such as structural changes due to cooperative Jahn-Teller (JT) distortions of Mn$^{3+}$ in octahedral environments, Mn migration, and phase transformations to spinel-like order, all of which affect the electrochemical performance. These physically complex phenomena motivate an \emph{ab initio} re-examination of the Li-Mn-O rock-salt space, with a focus on the thermodynamics of the prototypical \ch{LiMnO2} polymorphs. It is found that the generalized gradient approximation (GGA -- PBEsol) and meta-GGA (r$^2$SCAN) density functionals with empirically fitted on-site Hubbard $U$ corrections yield spurious stable phases for \ch{LiMnO2}, such as predicting a phase with $\gamma$-\ch{LiFeO2}-like order ($\gamma$-\ch{LiMnO2}) to be the ground state instead of the orthorhombic (Pmmn) phase, which is the experimentally known ground state. Accounting for antiferromagnetic order in each structure is shown to have a substantial effect on the total energies and resulting phase stability. By using hybrid-GGA (HSE06) and GGA with self-consistent Hubbard parameters (on-site $U$ and inter-site $V$) calculated from linear response theory, the experimentally observed \ch{LiMnO2} phase stability trends are recovered. The calculated on-site $U$ between Mn-$3d$ states in the experimentally observed orthorhombic, layered, and spinel phases are significantly smaller than $U$ in $\gamma$-\ch{LiMnO2} and disordered layered structures, by $0.5-0.6$ eV within GGA. The smaller values of $U$ are shown to be correlated with a collinear ordering of JT distortions, in which all $e_g$ orbitals are oriented in the same direction. This cooperative JT effect can lead to greater electron delocalization from Mn along the $e_g$ states due to increased Mn-O covalency, which contributes to the greater electronic stability compared to the phases with noncollinear JT arrangements. The structures with collinear ordering of JT distortions also generate greater vibrational entropy, which helps stabilize these phases at high temperature. These phases are shown to be strongly insulating with large calculated band gaps $>$ 3 eV, which are computed using HSE06 and $G_0W_0$.
\end{abstract}

\keywords{Mn-rich cathodes, electronic structure, first-principles calculations}

\maketitle
\clearpage

\section{\label{sec:intro}Introduction}
The manufacturing and deployment of lithium-ion (\ch{Li}) batteries have experienced rapid growth, driven by their application in electric vehicles, large-scale energy storage of intermittent energy sources (e.g. wind and solar power), and personal electronic devices \cite{goodenough-perspective-2013,Thackeray2012, Tian2020, Xie2021, maisel-lib-demand-2023}. To meet the growing demand for \ch{Li}-ion batteries, it is imperative that sustainable cathode chemistries based on earth-abundant elements are developed to prevent potential supply chain bottlenecks \cite{olivetti-supply-joule-2017, maisel-lib-demand-2023}. Manganese (\ch{Mn})-rich cathodes have emerged as strong candidates for such technologies, as \ch{Mn} is significantly cheaper and more earth-abundant than nickel (\ch{Ni}) and cobalt (\ch{Co}), which are the commonly used transition metals (TMs) in commercialized Li-ion batteries \cite{olivetti-supply-joule-2017, thackeray-lmo-1997}.

Many conventional cathode materials have been synthesized with the chemical formula \ch{Li$M$O2}, where $M$ can be either a single species or an admixture of TMs and other metals. These are rock-salt type structures, in which the oxygen (\ch{O}) sublattice is face-centered cubic (FCC) and the cations are distributed in the interstitial octahedral sites of the anion sublattice. Commonly used \ch{Li$M$O2} cathode compositions are typically rich in either Ni or Co, and crystallize in a layered structure with the R$\bar{3}$m space group \cite{manthiram-ni-rich-2016, goodenough-lco-1980, wu-limo2-philmag-1998}. \ch{LiMnO2} is particularly unique because its ground state is the orthorhombic (Pmmn) phase \cite{wu-limo2-philmag-1998, mishra-lmo-stability-prb-1998, dittrich-limno2-zaac-1969} -- in contrast with the Ni and Co counterparts -- and is characterized by corrugated layers of \ch{Li} and \ch{Mn}. This type of cation ordering is not observed as the ground state in any other reported \ch{Li$M$O2} composition \cite{wu-limo2-philmag-1998}.

\begin{figure*}
    \centering
    \includegraphics[scale=0.63]{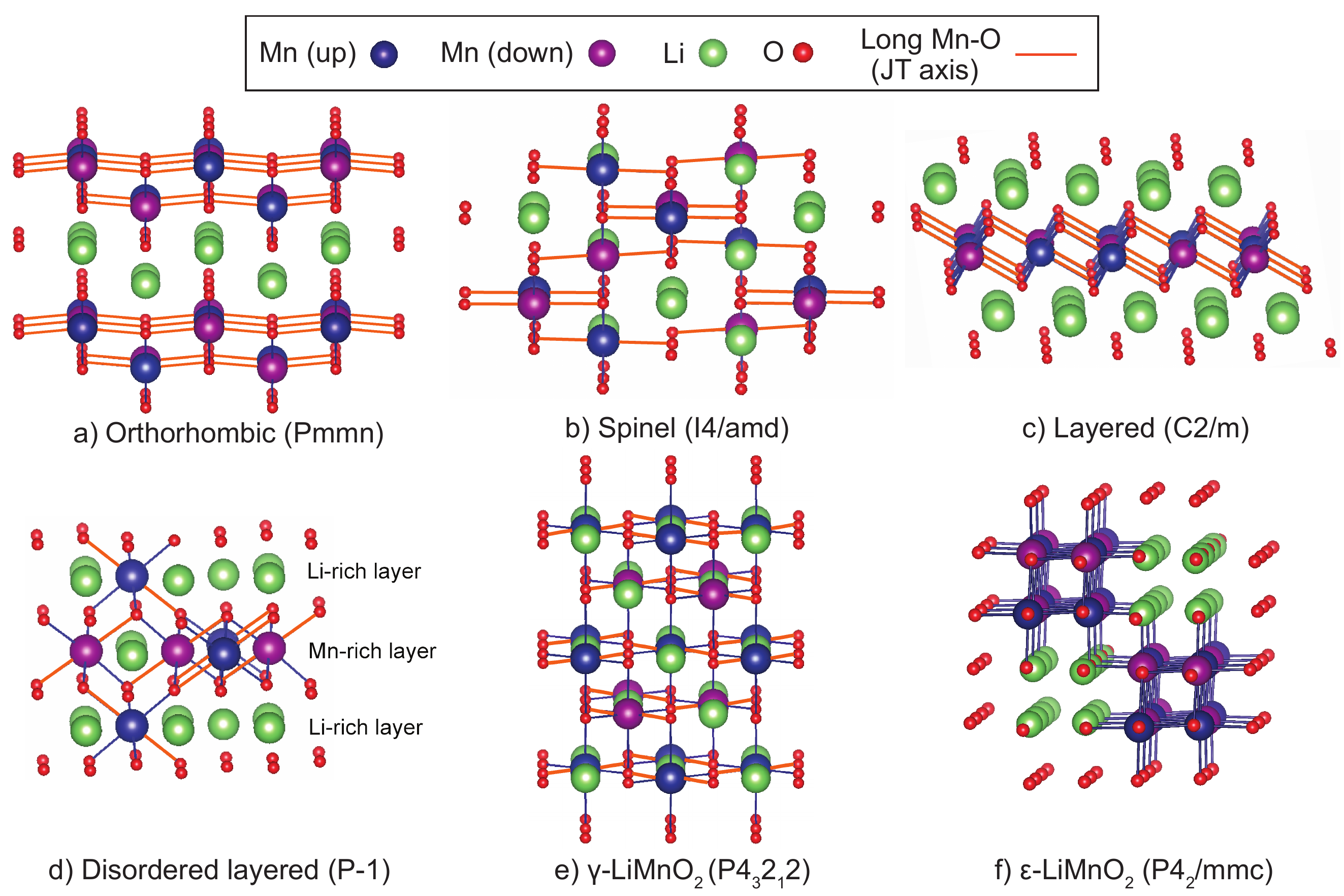}
    \caption{Structures of the \ch{LiMnO2} phases that are examined in this study. The majority (up, purple) and minority (down, blue) spin channels of \ch{Mn} are distinguished to display the ground state AFM orderings that we identify. The long Mn-O bonds (JT axis) are shown in orange and Li-O bonds are omitted for clarity. The space groups are labeled in parentheses. The spinel phase (b) is specifically the lithiated spinel (\ch{Li2Mn2O4} in spinel notation). In $\epsilon$-\ch{LiMnO2} (f), all JT axes are aligned perpendicular to the page and are not visible in the figure.}
    \label{fig:all_strucs}
\end{figure*}

To date, various \ch{LiMnO2} phases have been explored as potential cathode materials. In particular, three \ch{LiMnO2} polymorphs have been experimentally reported: (i) the aforementioned orthorhombic (Pmmn), (ii) the (lithiated) spinel (I4/amd), and (iii) the layered (C2/m) phases; these are shown in Figures \ref{fig:all_strucs}a-c, respectively \cite{dittrich-limno2-zaac-1969, armstrong-layered-limno2-nature-1996, capitaine-layered-limno2-ssi-1996, david-li2mn2o4-nd-jssc-1987}. In these \ch{LiMnO2} phases, a Jahn-Teller (JT) distortion of Mn$^{3+}$ occurs, characterized by the elongation of two axial \ch{Mn}-\ch{O} bonds and the shortening of four equatorial \ch{Mn}-\ch{O} bonds within the \ch{MnO6} octahedra \cite{kanamori-jt-1960}. The direction of the (longer) axial Mn-O bonds, which we will refer to as the JT axis, are highlighted in orange in Figure \ref{fig:all_strucs}. The JT axes within the 
experimentally refined orthorhombic (ortho), spinel, and layered \ch{LiMnO2} phases are collinearly aligned (Figures \ref{fig:all_strucs}a-c) \cite{greedan-ortho-nd-jssc-1997, hoppe_ortho_1975, akimoto_limno2_drx, armstrong-layered-limno2-nature-1996, capitaine-layered-limno2-ssi-1996, david-li2mn2o4-nd-jssc-1987}. The collinear ordering of JT distortions leads to a lowering of symmetry in layered \ch{LiMnO2} (C2/m space group) compared to layered \ch{LiNiO2} and \ch{LiCoO2} (R$\bar{3}$m space group) \cite{wu-limo2-philmag-1998, manthiram-ni-rich-2016, goodenough-lco-1980}. The tetragonal symmetry of spinel \ch{LiMnO2} (or \ch{Li2Mn2O4} in spinel notation) also arises from the collinear ordering of JT distortions, as the delithiated states of spinel \ch{LiMn2O4} and $\lambda$-\ch{MnO2}, which contain some or all Mn$^{4+}$, have cubic symmetry (Fd$\bar{3}$m) at ambient temperature \cite{david-li2mn2o4-nd-jssc-1987, thackeray-li-mn-spinels-1983}.

Previous experiments have indicated that these \ch{LiMnO2} phases are thermodynamically competitive with each other. The ortho phase is experimentally known to be the ground state \cite{greedan-ortho-nd-jssc-1997, dittrich-limno2-zaac-1969}, but differences in the synthesis precursors or conditions can lead to layered or spinel forming as impurity phases \cite{reimers-lt-limno2-jecs-1993, gummow-lt-ortho-mrb-1993, gummow-limno2-synth-jecs-1994}. The substitution of some Mn by \ch{Al} or many other metals can lead to increased stability of the layered phase \cite{jang-al-limno2-ssi-2000, mishra-lmo-stability-prb-1998}. Phase-pure layered \ch{LiMnO2} can be synthesized typically through ion-exchange from \ch{NaMnO2} \cite{armstrong-layered-limno2-nature-1996, capitaine-layered-limno2-ssi-1996}. Phase-pure spinel \ch{LiMnO2} can be formed by lithiating spinel \ch{LiMn2O4} \cite{thackeray-li-mn-spinels-1983}. When layered or ortho-\ch{LiMnO2} are electrochemically cycled, they tend to transform to a spinel-like phase \cite{armstrong-lay-spin-chem-mat-2004, reimers-lt-limno2-jecs-1993, gummow-lt-ortho-mrb-1993, gummow-limno2-synth-jecs-1994}. 

These experimental observations suggest that the free energy differences among these \ch{LiMnO2} phases should be relatively small. Thus, quantifying the relative stability of these polymorphs using \textit{ab initio} calculations would provide valuable insights, and is the objective of our investigation. As we will show, capturing the relative energy differences among \ch{LiMnO2} polymorphs is complicated by the strong interplay between electron localization, JT distortions, and magnetic ordering, making this a challenging test case for modern electronic structure theory. In an early study, Mishra et al. showed that capturing antiferromagnetic (AFM) order and use of accurate density-functional theory (DFT) techniques are required to predict ortho as the lowest energy structure among the experimentally known \ch{LiMnO2} phases \cite{mishra-lmo-stability-prb-1998}. In our study we apply a similar analysis, but using more recently developed DFT methodologies. We further investigate additional \ch{LiMnO2} orderings that have not been experimentally reported and previously overlooked in \emph{ab initio} studies, but that we find to have comparable energies to the known phases. These phases are: i) the disordered layered phase, which is a layered structure containing 25\% mixing between \ch{Li} and \ch{Mn} layers, ii) $\gamma$-\ch{LiMnO2}, which has the same cation order as the $\gamma$-\ch{LiFeO2} phase \cite{barre-nd-lifeo2-2009}, and iii) $\epsilon$-\ch{LiMnO2} where \ch{Mn} and \ch{Li} are ordered into nearest cation neighbor \ch{Mn4} and \ch{Li4} tetrahedral columns. The $\epsilon$ phase is a low-energy structure that we have found from Monte Carlo simulations performed using a cluster expansion lattice model trained on the rock-salt \ch{LiMnO2}-\ch{Li2MnO3} phase space (more details in Supplementary Information (SI) section VI \cite{si}). The disordered layered, $\gamma$, and $\epsilon$-\ch{LiMnO2} phases are shown in Figure \ref{fig:all_strucs}d-f, respectively. We note that the (DFT-relaxed) structures of disordered layered (Figure \ref{fig:all_strucs}d) and $\gamma$ (Figure \ref{fig:all_strucs}e) contain JT axes that are not all collinearly aligned, which is different from ortho, layered, spinel, and $\epsilon$ (Figures \ref{fig:all_strucs}a,b,c,f).

In order to accurately model these \ch{LiMnO2} phases, \emph{ab initio} methodologies that can precisely capture the localized nature of Mn-$3d$ states are needed, which presents a challenge for DFT approximations \cite{anisimov-ldau-1991, anisimov-ldau-cond-mat-1997}. Specifically, the self-interaction error (SIE) \cite{pz-sic-1981} that is inherent to semi-local DFT approximations tends to be especially large in localized $3d$ states, and it can contribute to the over-delocalization of $3d$ electrons, incorrect prediction of metallic states in insulating TM oxides \cite{anisimov-ldau-cond-mat-1997, dudarev-ldau-prb-1998}, and poor prediction of TM oxidation reaction energies \cite{wang-oxidation-ggau-prb-2006}. To overcome these limitations, Hubbard corrections are conventionally added to generalized gradient approximation (GGA) functionals \cite{wang-oxidation-ggau-prb-2006, zhou-redox-ldau-2004}. It has been argued early on that the key role of this Hubbard $U$ term is to enforce piecewise linearity of the Hubbard manifold, resulting in the correction of SIE \cite{cococcioni-ldau-prb-2004, kulik_dftu_2006}. Within this framework, the on-site Hubbard $U$ corrective term physically translates to a local potential constraining the electronic states projected onto the TM-$3d$ manifold to be as spatially confined as possible \cite{anisimov-ldau-1991, dudarev-ldau-prb-1998, Anisimov1993, cococcioni-ldau-prb-2004}. In our calculations, the Hubbard $U$ that we impose is on the Mn-$3d$ manifolds, unless otherwise stated.

Besides purely on-site contributions, it has been shown that applying an additional inter-site Hubbard $V$ parameter ($\mathrm{GGA}+U+V$) -- i.e. a hybridizing interaction between TM-$3d$ and O-$2p$ nearest neighbors -- can further improve predictions of structural, energetic, and magnetic properties in TM oxides exhibiting strong TM-O covalency \cite{cococcioni-ggauv-2010, timrov-olivine-prx-2022, binci-nickelates-prr-2023, mahajan-mno2-prm-2021, TancogneDejean2018, Jang2023}. The Hubbard parameters (HP) are generally determined either by empirical fitting of experimentally known properties (e.g. band gaps, oxidation enthalpies) \cite{wang-oxidation-ggau-prb-2006, gautam-scanu-prm-2018, swathilakshmi-r2scanu-jctc-2023}, or from \textit{ab initio} calculations, which can be based on linear response theory DFT \cite{Anisimov1991, cococcioni-ldau-prb-2004,timrov-hp-dfpt-prb-2018,binci_noncoll_2023,moore-ggauj-prm-2024}, constrained random-phase approximation \cite{Kotani2000, Aryasetiawan2004,Aryasetiawan2006}, Hartree-Fock based methodologies \cite{Mosey2007,Agapito2015,TancogneDejean2018,Lee2020} and machine learning techniques \cite{Yu:2020,Cai:2024,uhrin:2024}.
When HP are obtained from empirical fitting, they are generally held constant for each TM element across all the considered structures. This scheme is frequently used, especially for high throughput workflows, as the computationally expensive calculation of HP can be avoided \cite{wang-oxidation-ggau-prb-2006}. However, the \textit{ab initio} calculated HP for a given species can vary significantly depending on the local environment and oxidation state \cite{timrov-hp-dfpt-prb-2018, timrov-spinel-pccp-2023, moore-ggauj-prm-2024} -- variations that cannot be captured by empirical fitting procedures.

Another way of building upon the GGA functional is to incorporate electron kinetic-energy density corrections to the exchange-correlation energies, in the so-called meta-GGA functionals. Meta-GGAs, particularly the SCAN family (SCAN, rSCAN, r$^2$SCAN), have consistently outperformed GGAs in predicting formation enthalpies and structural properties \cite{sun-scan-prl-2015, sun_accurate_2016, bartok-r2scan-2019, furness-r2scan-2020, kingsbury-r2scan-bench-prm-2022, hinuma-dft-oxides-2017, yang_rationalizing_2019}. Kitchaev et al. further demonstrated SCAN's superior accuracy in modeling the \ch{MnO2} polymorph energetics within TM oxides \cite{kitchaev-mno2-prb-2016}. Given the improved performance of SCAN and r$^2$SCAN over GGA, Hubbard corrections are not commonly applied to these functionals, although Gautam et al. recently demonstrated that adding an empirically fitted Hubbard $U$ correction to SCAN and r$^2$SCAN can enhance the prediction of oxidation enthalpies and polymorph ground states in several TM oxides \cite{gautam-scanu-prm-2018, swathilakshmi-r2scanu-jctc-2023}.

An alternative strategy to mitigate the self-interaction error (SIE) in generalized gradient approximations (GGAs) involves incorporating a fraction of exact (Fock) exchange into the electronic exchange energy calculation, as implemented in hybrid functional methods \cite{Kummel2008, Paier2006}. Hybrid functionals, particularly HSE06 \cite{heyd-screen-hybrid-2003}, have been demonstrated to accurately predict lattice parameters, formation energies, and band gaps of TM oxides \cite{hse06-2006, chevrier-hse-tmo-2010, Toroker2011}. For a rigorous evaluation of quasiparticle band structures, however, the $GW$ method provides a more robust many-body framework, especially for the calculation of electron excitation energies \cite{huser-gw-solids-2013, ergonenc-gw-tmo-2018}. While $GW$ generally performs well for predicting band gaps of conventional semiconductors \cite{Hybertsen1985, Shishkin2007, Schilfgaarde2006}, a careful selection of the DFT starting point has to be made when dealing with TM oxides, as the accuracy of the $GW$ self-energy evaluation heavily depends on the quality of the electronic ground state, especially for single-shot ($G_0W_0$) calculations \cite{Hong2010, Rodl2009}.

In this work, we apply the DFT methods we discussed above, namely GGA+$U$(+$V$), r$^2$SCAN(+$U$), and HSE06 to evaluate the phase stability of the \ch{LiMnO2} polymorphs. The paper is organized as follows: in Section \ref{sec:methods}, we provide the details of the computational schemes we used, while in Section \ref{sec:phase_stab}, the \ch{LiMnO2} phase stability results are presented. We find that accounting for AFM order is vital to obtain more accurate phase stability trends across several levels of theory, which is consistent with the results of Mishra et al. \cite{mishra-lmo-stability-prb-1998} and Singh \cite{Singh1994}. We encounter challenges when applying PBEsol \cite{pbe-sol-2008} and r$^2$SCAN \cite{furness-r2scan-2020} with or without empirically tuned values of Hubbard $U$, to the various AFM \ch{LiMnO2} phases, as the predicted ground state is the $\gamma$-\ch{LiMnO2} phase and not the orthorhombic \ch{LiMnO2}, which is inconsistent with experimental observations. We are thus required to assess more computationally complex schemes such as HSE06 and PBEsol with self-consistent Hubbard parameters. Using these computational methodologies, we recover relative energies in closer agreement with experiments. We assess the phase stability at elevated temperature by performing harmonic phonon calculations within $\mathrm{PBEsol}+U$ and determine the vibrational free energy as a function of temperature. To rationalize the differences in predicted phase stability, we analyze the properties related to the electronic structure that each method yields. Specifically, in Section \ref{sec:sc_hubbard}, the self-consistently calculated HP in all \ch{LiMnO2} phases and their correlation with the structural and electronic properties are examined. In Section \ref{sec:el_structure}, we analyze electronic properties such as the electron density, density of states, and band structure of ortho-\ch{LiMnO2} across the different DFT approximations. In the absence of experimental data on the electronic structure, we present theoretical predictions of the band structure and excitation energies. Our calculations using HSE06 and $G_0W_0$ characterize the system as a wide band gap insulator, with the band gap estimated to be $>$ 3 eV. In the Discussion (Section \ref{sec:discussion}), we further rationalize the relation between the ion configuration, electronic structure, and phase stability trends in these polymorphs.

\section{\label{sec:methods}Methods}

We use a variety of DFT functionals to perform structural relaxations and calculate total energies of the \ch{LiMnO2} phases. The Vienna \textit{ab initio} Simulation Package (VASP) \cite{kresse1993,kresse1994,kresse_efficient_1996,kresse1996a} is used for the structural relaxations performed within $\mathrm{PBEsol}+U$, r$^2$SCAN(+$U$), and HSE06. When "$+U$" is used, a Hubbard $U$ is applied on the Mn-$3d$ manifolds with values of $U$ that have been empirically fitted in previous studies \cite{wang-oxidation-ggau-prb-2006,swathilakshmi-r2scanu-jctc-2023}. The \texttt{pymatgen} package is used to generate and manipulate structures, create calculation inputs, parse the calculation outputs \cite{ong_pymatgen_2013}, identify symmetries, and compute X-ray diffraction (XRD) patterns. \texttt{Quantum ESPRESSO} (\texttt{QE}) \cite{qe-2009, qe-2017} is used for its implementation of extended Hubbard parameters ($\mathrm{PBEsol}+U+V$) and density functional perturbation theory (DFPT) for self-consistent calculations of Hubbard $U$ and $V$ parameters \cite{timrov-hp-dfpt-prb-2018,timrov-self-prb-2021,timrov-hp-2022,timrov2020pulay}. The details of the DFT calculations, including the pseudopotentials, energy cutoff, and convergence criteria are listed in Supplementary Information (SI) Section I \cite{si}. To demonstrate the consistency between the energy differences obtained with VASP and QE, we use them to evaluate the energies of the \ch{LiMnO2} phases using $\mathrm{PBEsol}+U$, scanning $U$ from 3 to 7 eV. The results are reported in SI Figure S1 \cite{si}, which show good compatibility between the two codes.

The $\epsilon$ phase was found from canonical Monte Carlo (MC) simulations at the \ch{LiMnO2} composition, using a cluster expansion (CE) on a rock-salt Li$^{+}$-Mn$^{3+}$-Mn$^{4+}$-O$^{2-}$ lattice. The CE model construction and MC simulations were performed using \texttt{smol} \cite{smol}. More details are described in SI Section VI \cite{si}.

The AFM ground state of each phase is identified by enumerating $\sim$ 30 collinear AFM orderings and relaxing each structure within $\mathrm{PBEsol}+U$ ($U$ = 3.9 eV, VASP). The AFM orderings were enumerated using the \texttt{MagneticStructureEnumerator} module developed by Horton et al. in \texttt{pymatgen} \cite{horton-afm-search-2019, ong_pymatgen_2013}. The AFM ground state of ortho-\ch{LiMnO2} that we identify is identical to the AFM structure that was experimentally refined by Greedan et al. using neutron diffraction (ND) \cite{greedan-ortho-nd-jssc-1997}.

The self-consistent calculations of the Hubbard parameters (HP) within PBEsol are performed using DFPT as implemented in the \texttt{HP} module of \texttt{QE} \cite{timrov-hp-2022}. To converge the HP, we perform iterations in which $U$ and $V$ are computed using DFPT followed by a relaxation of the structure with these HP within $\mathrm{PBEsol}+U+V$. This procedure is repeated until all $U$ and $V$ values are converged to within 10 meV between iterations. The values of the converged HP are shown and analyzed in Section \ref{sec:sc_hubbard}. Each DFPT calculation requires the ground state electron density and wavefunctions, which are computed from static self-consistent field (SCF) calculations within $\mathrm{PBEsol}+U+V$. We also compute the self-consistent Hubbard $U$ within r$^2$SCAN using \texttt{QE}, using the finite-difference linear response approach based on supercells \cite{cococcioni-ldau-prb-2004} instead of DFPT, since the current version of \texttt{QE} (version 7.4) does not support DFPT calculations with meta-GGA functionals. The \rrscan calculations in \texttt{QE} are performed using the interface with the \texttt{Libxc} library of density functionals \cite{libxc}. Furthermore, we perform the self-consistent calculation of $U$ for \rrscan at the structural geometries relaxed using VASP, since variable-cell structural optimizations are not yet implemented for \rrscan in \texttt{QE}. Due to the current limitations of the codes, we present the \rrscan + $U_\mathrm{sc}$ results in the SI. More details of the self-consistent HP calculations are shown in SI Section II \cite{si}.

To obtain the vibrational properties, harmonic phonon calculations are computed using the frozen phonon method with Phonopy and VASP \cite{togo_phonopy_2015, kresse_efficient_1996}. Each \ch{LiMnO2} phase is relaxed within its respective AFM ground state and atomic displacements are generated on a supercell. The dynamical matrix is constructed by computing the forces in each supercell by performing SCF calculations. We use $\mathrm{PBEsol}+U$ ($U$ = 3.9 eV) for the phonon calculations, as it has been previously shown that $\mathrm{PBEsol}+U$ can predict phonon frequencies with good accuracy in a range of TM oxides \cite{zhang-cuprate-epc-2007, floris-mno-nio-dftu-2011, floris-dfpt-usp-2020, petretto-dfpt-workflow-2018, floris-dfpt-usp-2020}.

In Section \ref{sec:el_structure}, we compute the electronic structure of ortho-\ch{LiMnO2} using the range of DFT functionals we have discussed, to obtain the electron density, density of states, magnetic moments, and band gaps. For these calculations, we fix the atomic positions to the primitive cell of the AFM ortho-\ch{LiMnO2} structure that was previously refined from ND by Greedan et al. \cite{greedan-ortho-nd-jssc-1997} The band structure of ortho-\ch{LiMnO2} is computed using $\mathrm{PBEsol}+U$, HSE06, and single-shot $GW$ ($G_0W_0$). SeeK-path is used to select the high symmetry path through the first Brillouin zone \cite{seekpath-2017}, which satisfies the magnetic symmetry of the primitive cell of AFM ortho-\ch{LiMnO2} described by the C2/c space group, which is of lower symmetry than the non-magnetic structure (Pmmn). The $\mathrm{PBEsol}+U$ band structure is computed using QE and its ground state wave functions are used as a starting point for the $G_0W_0$ calculations, which are performed using BerkeleyGW \cite{hybertsen-gw-1986, bgw-2012}. The single-particle Green's function ($G$) and screened Coulomb interaction ($W$) are constructed using the wave functions generated from $\mathrm{PBEsol}+U$. To calculate $W$, the dielectric matrix ($\epsilon$) is computed within the random phase approximation. More details are shown in SI Section V \cite{si}.

\section{\label{sec:results}Results}
\subsection{\label{sec:phase_stab}\ch{LiMnO2} Phase Stability}
\begin{figure}[!h]
    \begin{subfigure}{0.48\textwidth}
        \includegraphics[scale=0.51]{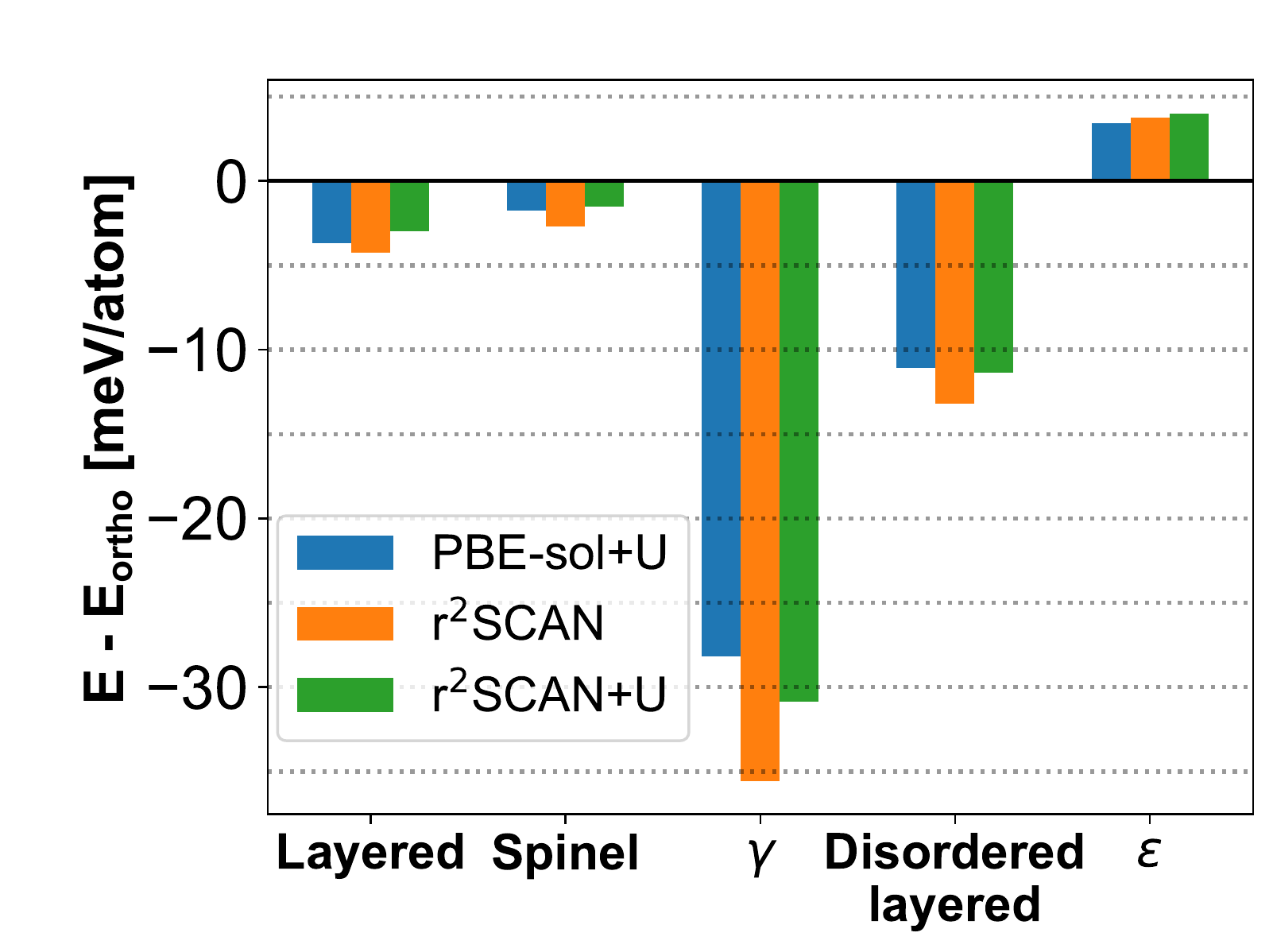}
        \caption{Ferromagnetic (FM)}
        \label{fig:fm_gs_ens}
    \end{subfigure}
    \begin{subfigure}{0.48\textwidth}
        \includegraphics[scale=0.51]{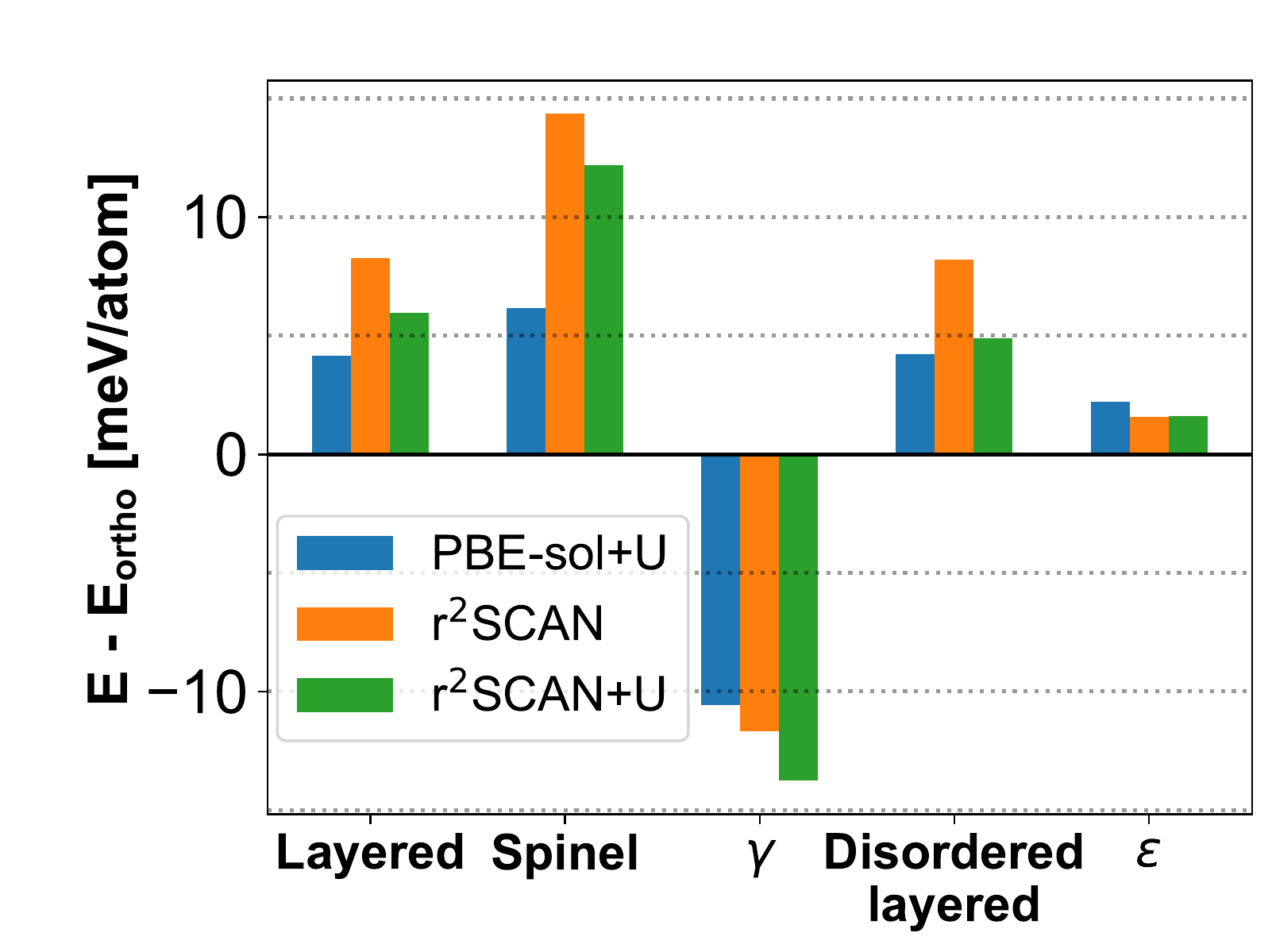}
        \caption{Antiferromagnetic (AFM)}
        \label{fig:afm_gs_ens}
    \end{subfigure}
    \begin{subfigure}{0.48\textwidth}
        \includegraphics[scale=0.51]{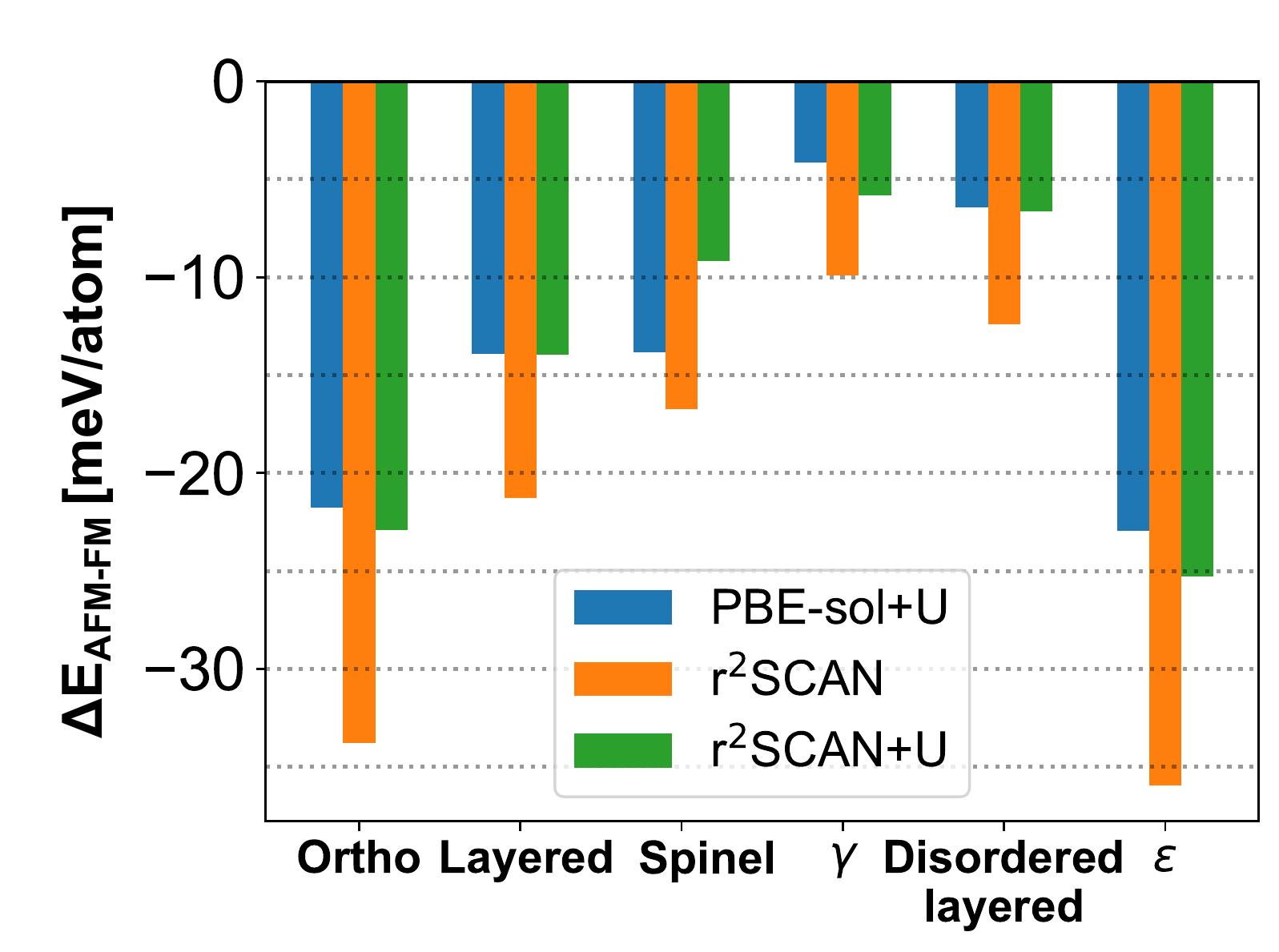}
        \caption{AFM-FM energy difference}
        \label{fig:afm_fm_en_diff}
    \end{subfigure}

    \caption{DFT energy of \ch{LiMnO2} phases relative to ortho-\ch{LiMnO2} in the a) FM and b) AFM spin configurations, and c) energy differences between the AFM and FM spin orderings ($\Delta E_{AFM-FM}$). Energies are computed with $\mathrm{PBEsol}+U$ ($U$ = 3.9 eV), r$^2$SCAN, and r$^2$SCAN+U ($U$ = 1.8 eV) in VASP. }
    \label{fig:ggau_r2scan_gs_ens}
\end{figure}

We evaluate the phase stability of the \ch{LiMnO2} polymorphs at 0 K by performing structural relaxations and calculating the total energy using $\mathrm{PBEsol}+U$, \rrscan, and \rrscan+ $U$ ($U$(PBEsol) = 3.9 eV \cite{wang-oxidation-ggau-prb-2006} and $U$(\rrscan) = 1.8 eV \cite{swathilakshmi-r2scanu-jctc-2023}). Within each functional, we evaluate the energy of each structure relative to ortho-\ch{LiMnO2} with all structures containing either FM or AFM spin ordering, as shown in Figures \ref{fig:fm_gs_ens} and \ref{fig:afm_gs_ens}, respectively. Hereafter, we use the notation $\Delta E_\mathrm{X}\equiv E_\mathrm{X}-E_\mathrm{ortho}$ to represent the energy of phase X relative to ortho-\ch{LiMnO2}, where a positive value of $\Delta E_\mathrm{X}$ implies that ortho-\ch{LiMnO2} is lower in energy than X.

The results in Figure \ref{fig:fm_gs_ens} show that with FM spin order, the ortho phase is higher in energy than \emph{all} the considered \ch{LiMnO2} polymorphs, with the exception of $\epsilon$. The $\gamma$-\ch{LiMnO2} structure is clearly the ground state as $\Delta E_{\gamma} \approx -30$ meV/atom. Disordered layered is the second lowest in energy and is $\sim$ 8 meV/atom below layered \ch{LiMnO2}. Layered and spinel are both predicted to be slightly lower in energy than ortho by $<$ 5 meV/atom. The trends produced by all the applied DFT functionals are similar.

These results clearly contradict experimental observations reporting ortho as the ground state \cite{greedan-ortho-nd-jssc-1997, dittrich-limno2-zaac-1969}. The prediction of $\gamma$ to be the ground state is almost certainly incorrect, as the $\gamma$-\ch{LiFeO2}-like cation ordering has never been reported for the \ch{LiMnO2} composition. Furthermore, disordered layered is predicted to be lower in energy than layered, implying a negative anti-site defect formation energy, which is inconsistent with the synthesis of well-ordered metastable layered \ch{LiMnO2} \cite{armstrong-layered-limno2-nature-1996, capitaine-layered-limno2-ssi-1996}.

With AFM spin ordering (Figure \ref{fig:afm_gs_ens}), $\gamma$ remains the ground state, but its energetic advantage over the orthorhombic structure has decreased compared to the FM case (Figure \ref{fig:fm_gs_ens}). The $\epsilon$ structure is again slightly higher in energy than ortho but more stable than layered and spinel. There is no significant energy difference between disordered layered and layered \ch{LiMnO2}.

The notable differences in phase stability predicted with AFM order compared to the FM case indicate that the influence of magnetic order on the total energy varies greatly by phase, consistent with earlier findings \cite{mishra-lmo-stability-prb-1998}. To quantify the effect of magnetic configuration, we calculate the energy difference between the AFM and FM configurations ($\Delta E_\mathrm{AFM-FM}$) for each \ch{LiMnO2} phase and plot them in Figure \ref{fig:afm_fm_en_diff}. We find that all structures are significantly more stable in their AFM ground state compared to the FM state. Ortho and $\epsilon$ are stabilized the most by AFM order, with $\Delta E_\mathrm{AFM-FM}$ $\approx$ $-20$ to $-25$ meV/atom ($\approx$ $-80$ to $-100$ meV/Mn) within $\mathrm{PBEsol}+U$ or \rrscan+ $U$. While AFM order also lowers the energy of layered and spinel, ($\Delta E_\mathrm{AFM-FM}$ $\approx$ -10 to -15 meV/atom) its effect is slightly smaller in magnitude than for ortho and $\epsilon$. Within the $\gamma$ and disordered layered structures, $\Delta E_\mathrm{AFM-FM}$ is smaller ($\sim$ 5 meV/atom). These variations in $\Delta E_\mathrm{AFM-FM}$ explain why the order of phase stability changes when all structures are in their AFM ground state compared to the FM case. We note that the absolute values of $\Delta E_\mathrm{AFM-FM}$ calculated from \rrscan are consistently larger compared to $\mathrm{PBEsol}+U$ and \rrscan+ $U$. Despite this discrepancy, the qualitative trends of $\Delta E_\mathrm{AFM-FM}$ among these phases are similar across these DFT approximations.

The phase stability predicted within AFM order are more consistent with experiment than in the FM case, as ortho is lower in energy than layered and spinel (Figure \ref{fig:afm_gs_ens}). But certain trends are still questionable, as $\gamma$ is the predicted ground state and $\epsilon$ is lower in energy than layered and spinel. Furthermore, the degeneracy of disordered layered and layered would indicate that there is no energy requirement to form specific anti-site defects in layered. We thus assess the phase stability of the \ch{LiMnO2} polymorphs using electronic structure methods that enforce electron localization in more select ways, such as $\mathrm{PBEsol}+U+V$ and the HSE06 hybrid functional. We also assess whether employing self-consistently calculated $U$ and $V$ parameters ($\mathrm{PBEsol}+(U+V)_\mathrm{sc}$) can significantly affect the results. We thus evaluate the energy of each AFM \ch{LiMnO2} phase relative to ortho using HSE06, $\mathrm{PBEsol}+U_\mathrm{sc}$, and $\mathrm{PBEsol}+(U+V)_\mathrm{sc}$, with the results shown in Figure \ref{fig:dft_uv_hse_gs_ens}.

\begin{figure}
    \centering
    \includegraphics[scale=0.56]{250312_dft_ens_hse_dftuv.pdf}
    \caption{DFT energy of the \ch{LiMnO2} polymorphs relative to ortho, in their respective ground state AFM spin configurations. The functionals used are HSE06, PBEsol (GGA) with self-consistently (sc) determined $U$ (PBEsol+$U_\mathrm{sc}$), PBEsol+$(U+V)_\mathrm{sc}$, and PBEsol with averaged (avg) $U$ = 6 eV and $V$ = 0.6 eV.}
    \label{fig:dft_uv_hse_gs_ens}
\end{figure}

Using these DFT functionals, we largely recover the experimental trends across all the investigated \ch{LiMnO2} phases. HSE06 (purple bars in Figure \ref{fig:dft_uv_hse_gs_ens}) correctly predicts the ortho phase to be the ground state of \ch{LiMnO2}, although it is only marginally ($\approx1$ meV/atom) lower in energy than $\gamma$. The $\epsilon$ structure is only slightly higher in energy than ortho ($\Delta E_\epsilon\approx2$ meV/atom) and lower in energy than layered and spinel. Disordered layered becomes $\sim$ 7 meV/atom higher in energy than layered, which corresponds to a substantial anti-site defect formation energy ($\Delta E_\mathrm{a-s,f}$) of 112 meV/defect (there is one defect per \ch{Li4Mn4O8} in disordered layered).

The phase stability predicted within $\mathrm{PBEsol}+(U+V)_{\mathrm{sc}}$ (green bars in Figure \ref{fig:dft_uv_hse_gs_ens}) shares similar features with HSE06. Ortho is predicted to be the ground state. However, $\gamma$ and disordered layered are both significantly higher in energy relative to ortho ($\Delta E_\gamma\approx\Delta E_\mathrm{dis \ lay}\approx 25$ meV/atom). This energy difference corresponds to $\Delta E_\mathrm{a-s,f}\approx400$ meV/defect in layered, which is much larger than the value predicted by HSE06. The energy of layered, spinel, and $\epsilon$ relative to ortho-\ch{LiMnO2} are in close agreement with HSE06. If we neglect the intersite V parameter and include only the self-consistent on-site $U$ correction ($\mathrm{PBEsol}+U_\mathrm{sc}$ -- light blue bars in Figure \ref{fig:dft_uv_hse_gs_ens}), we find that the energy differences are very close to the $\mathrm{PBEsol}+(U+V)_{\mathrm{sc}}$ results. The only significant differences are that $\Delta E_\gamma$ and $\Delta E_\mathrm{dis \ lay}$ are slightly smaller in $\mathrm{PBEsol}+U_\mathrm{sc}$ compared to $\mathrm{PBEsol}+(U+V)_{\mathrm{sc}}$. Thus, capturing the variations in self-consistent $U$ gives the leading contribution for an accurate prediction of \ch{LiMnO2} phase stability.

We note that we have evaluated the \ch{LiMnO2} energetics using \rrscan+ $U_\mathrm{sc}$, as shown in SI Figure S3 \cite{si}, which largely recovers the same trends as $\mathrm{PBEsol}+U_\mathrm{sc}(+V_\mathrm{sc})$ -- namely ortho is the ground state, $\gamma$ is significantly higher in energy than ortho, and disordered layered is higher in energy than layered. There are discrepancies in our linear response calculations of $U$ within \rrscan compared to PBEsol, primarily 
since DFPT and variable-cell structural relaxations within \rrscan are not yet implemented in the current version of \texttt{QE}; for this reason these results are reported in the SI \cite{si}.

The \ch{LiMnO2} energies are also evaluated using PBEsol with averaged Hubbard parameters ($\mathrm{PBEsol}+(U+V)_\mathrm{avg}$) of $U$ = 6 eV and $V$ = 0.6 eV, which are roughly the average of the self-consistent $U$ and $V$ values across all the structures (red bars in Figure \ref{fig:dft_uv_hse_gs_ens}). $\mathrm{PBEsol}+(U+V)_\mathrm{avg}$ yields the spurious phase stability trends that were observed in $\mathrm{PBEsol}+U$ and r$^2$SCAN($+U$) (Figure \ref{fig:afm_gs_ens}) -- namely, $\gamma$ is the predicted ground state and disordered layered is nearly equal in energy to layered. Therefore, the self-consistently calculated and structurally-informed Hubbard parameters are essential in order to obtain relative phase stability that is more consistent with experiments.

\begin{figure}[!h]
    \centering
    \includegraphics[scale=0.54]{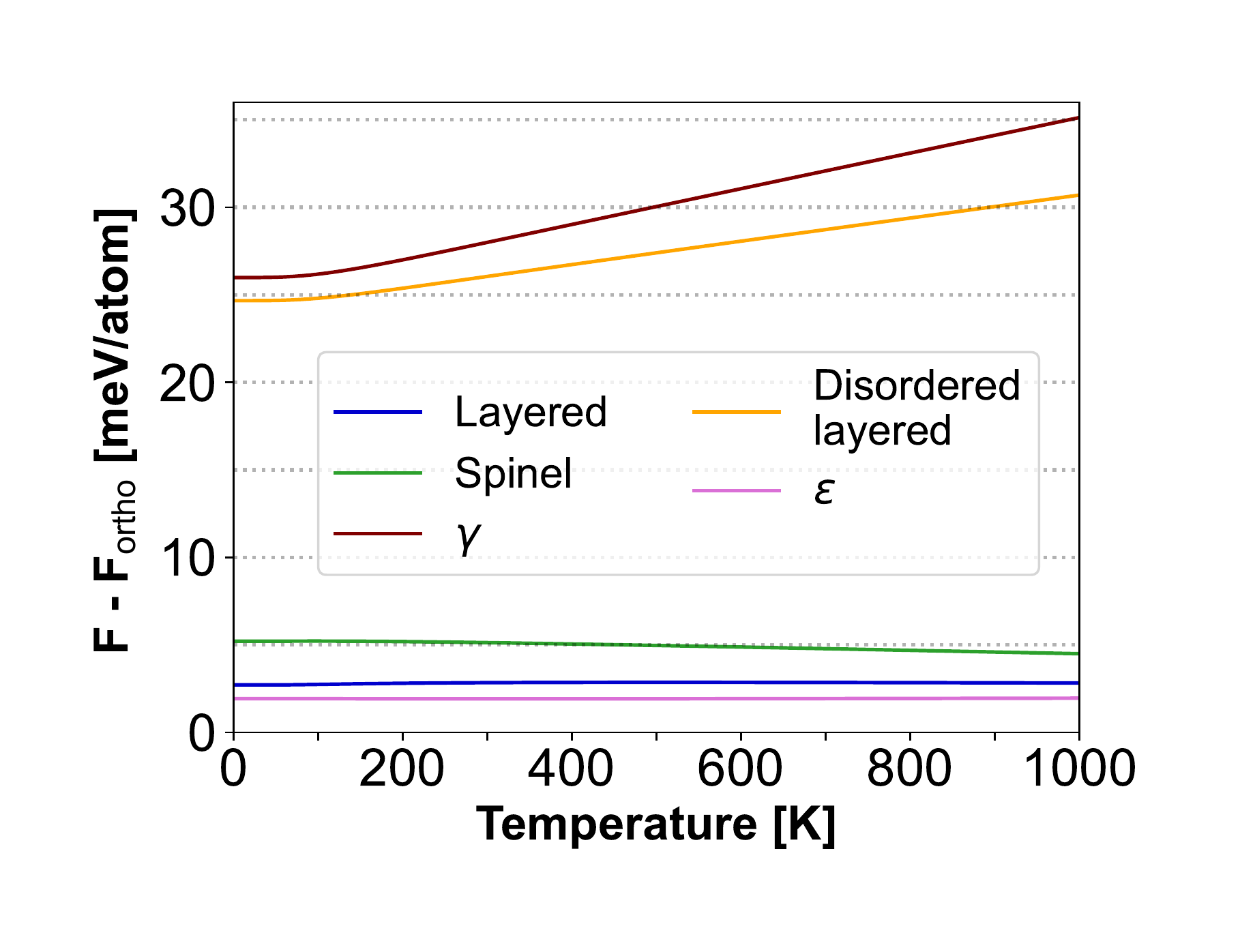}
    \caption{Total free energy (F) relative to ortho-\ch{LiMnO2} as a function of temperature. F(T) is the sum of the vibrational free energy and DFT energy. Harmonic phonon calculations are performed within $\mathrm{PBEsol}+U$ ($U$ = 3.9 eV) and DFT energies are computed within $\mathrm{PBEsol}+(U+V)_\mathrm{sc}$.}
    \label{fig:tot_f_energy}
\end{figure}

The phase stability assessed thus far has been restricted to $T=0$ K. We include finite-temperature effects by evaluating the harmonic phonon contribution to the free energy ($F_\mathrm{vib}(T)=E_\mathrm{vib}(T)-TS_\mathrm{vib}(T)$) and adding it to the electronic PBEsol+($U+V)_\mathrm{sc}$ total energy at 0 K ($E_\mathrm{ele})$: $F(T)=E_\mathrm{ele}(T = 0)+F_\mathrm{vib}(T)$. The calculated phonon density of states for each phase is shown in SI Figure S5 \cite{si}. For simplicity, we evaluate $F_\mathrm{vib}(T)$ only at the $\mathrm{PBEsol}+U$ level in the AFM configuration. 
Figure \ref{fig:tot_f_energy} shows the calculated $F(T)$ of each phase relative to ortho-\ch{LiMnO2} as a function of temperature from 0 to 1000 K (we denote $\Delta F_\mathrm{x}(T)$ as the free energy difference of phase X relative to ortho). The free energy of ortho is lower than all the other phases across the entire temperature range considered. The experimentally observed metastable layered and spinel are only $<$ 5 meV/atom higher in energy than ortho-\ch{LiMnO2} across this temperature range.
The $\epsilon$ phase is $\approx2$ meV/atom higher in energy than ortho and slightly lower in energy than both layered and spinel. The changes in $\Delta F_x(T)$ for layered, spinel, and $\epsilon$ are small ($<$ 1 meV/atom) across this temperature range. For the $\gamma$ and disordered layered phases, $\Delta F_x(T)$ increases significantly with temperature. Thus, the vibrational entropy appears to help stabilize ortho, layered, spinel, and $\epsilon$ relative to the $\gamma$ and disordered layered phases at elevated temperature.

\subsection{\label{sec:sc_hubbard}Self-consistent Hubbard parameters of \ch{LiMnO2}}

We have shown that the phase stability predicted with HSE06, $\mathrm{PBEsol}+U_\mathrm{sc}(+V_\mathrm{sc})$, and \rrscan + $U_\mathrm{sc}$ better reflects experimental observations, compared to those obtained with PBEsol and r$^2$SCAN, with or without empirical Hubbard parameters (HPs). Specifically, all the methods with empirical or averaged HPs, whether with an on-site $U$ or including inter-site $V$, yield the $\gamma$ phase to be the ground state. We believe this result to be almost certainly incorrect since $\gamma$ has never been experimentally reported for \ch{LiMnO2}. 

The self-consistently determined Hubbard parameters from the $\mathrm{PBEsol}+(U+V)_\mathrm{sc}$ and $\mathrm{PBEsol}+U_\mathrm{sc}$ calculations of each phase are listed in Table \ref{tab:hps}. Importantly, as shown by Timrov et al., these computed HPs are not transferable across different pseudopotentials or forms of Hubbard projectors, since they can have different degrees of localization \cite{timrov-hp-dfpt-prb-2018, timrov-hp-2022}. Within $\mathrm{PBEsol}+(U+V)_\mathrm{sc}$, the self-consistent $U$ of $\gamma$-\ch{LiMnO2} is 6.4 eV, which is $\sim$ 0.6 eV larger than that in ortho, layered, spinel, and $\epsilon$-\ch{LiMnO2} ($U$ $\sim$ 5.8 eV). In the disordered layered phase, the calculated $U$ span a range of values ($5.92-6.34$ eV).
Within $\mathrm{PBEsol}+U_\mathrm{sc}$, the trends among the calculated Hubbard $U$ are qualitatively similar, with the key difference being that the Hubbard $U$ parameters of each phase are systematically larger by $\sim 0.17$ eV on average compared to $\mathrm{PBEsol}+(U+V)_\mathrm{sc}$. Within \rrscan, the values of $U$ are $\sim 4.5$ eV (SI Table S2 \cite{si}), which are systematically smaller by $> 1$ eV compared to PBEsol; though we note that the differences between the Hubbard $U$ calculated within \rrscan and PBEsol arise in part due to the different pseudopotentials used (more details in SI Section II \cite{si}). The larger values of the calculated Hubbard $U$ in $\gamma$ and disordered layered are very likely the main reason why these phases are predicted to be unstable relative to ortho-\ch{LiMnO2} within $\mathrm{PBEsol}+U_\mathrm{sc}(+V_\mathrm{sc})$ and \rrscan + $U_\mathrm{sc}$. Therefore, the phase stability trends can be dictated by the nature of electron-electron interactions, which are more genuinely captured by the HPs calculated from the linear response framework.

\begin{table*}
\caption{Hubbard on-site $U$ (\ch{Mn} $3d$), inter-site $V$ (\ch{Mn}$3d$ - \ch{O} $2p$) parameters, and JT bond ratios ($\frac{\mathrm{avg. long}}{\mathrm{avg. short}}$) in the \ch{LiMnO2} phases. Hubbard parameters (in units of eV) are self-consistently calculated using DFPT. Only the distinct values of each property are listed. $V$ of the short (1.9-2.0 \AA) and long (2.2-2.3 \AA) \ch{Mn}-\ch{O} bonds are shown in separate rows. Reference values of the JT bond ratios from experimentally refined structures are listed.}
    \centering
    \begin{adjustbox}{width=\textwidth}
    \begin{tabular}{c c c c c c c c}
    \hline 
         Method & Property & Orthorhombic & Layered & Spinel & $\gamma$ & Disordered & $\epsilon$ \\
         & & & & & & layered & \\
         \hline 
         PBEsol & $U$ & 5.81 & 5.79 & 5.81 & 6.36 & 5.92, 6.25, 6.34 & 5.81 \\
         + $U+V$ & $V$ (short) & 0.72, 0.59 & 0.65 & 0.66, 0.65 & 0.74 - 0.79 & 0.64 - 0.77 & 0.71, 0.59 \\ 
         & $V$ (long) & 0.38 & 0.30 & 0.31 & 0.57, 0.54 & 0.36 - 0.57 & 0.35\\
          & JT bond ratio & 1.189 & 1.203 & 1.200 & 1.152 & 1.156, 1.211, 1.138 & 1.190 \\
         \hline 
         PBEsol & $U$ & 6.01 & 6.01 & 6.02 & 6.49 & 6.08, 6.40, 6.44 & 6.01 \\
         + $U$ & JT bond ratio & 1.181 & 1.191 & 1.189 & 1.134  & 1.148, 1.189, 1.125 & 1.181 \\
         \hline
         Exp & JT bond ratio & 1.182, 1.192 \cite{greedan-ortho-nd-jssc-1997, hoppe_ortho_1975} & 1.201 \cite{armstrong-layered-limno2-nature-1996} & -- & -- & -- & -- \\
         & & 1.194 \cite{akimoto_limno2_drx} & & & & & \\
         \hline
    \end{tabular}
    \end{adjustbox}
    \label{tab:hps}
\end{table*}

We also observe variations in the extent of Mn JT distortions in each phase, which we measure by evaluating the ratio between the averaged long Mn-O bond length and averaged short Mn-O bond length (we will refer to this quantity as the JT bond ratio = $\frac{\text{avg. long}}{\text{avg. short}}$). Within $\mathrm{PBEsol}+(U+V)_\mathrm{sc}$, the predicted JT bond ratio of $\gamma$ is significantly smaller than that of ortho, layered, spinel, and $\epsilon$ (Table \ref{tab:hps}). Within disordered layered, the JT bond ratios, as well as the Hubbard $U$, span a range of small and large values. These trends are also observed within $\mathrm{PBEsol}+U_\mathrm{sc}$, which suggests that the self-consistent $U$ and extent of JT distortion are related. The predicted JT bond ratios within $\mathrm{PBEsol}+U_\mathrm{sc}$ are systematically smaller compared to $\mathrm{PBEsol}+(U+V)_\mathrm{sc}$, which we attribute to the increased electron localization when excluding the inter-site $V$ term. In general, the JT bond ratios predicted by $\mathrm{PBEsol}+(U+V)_\mathrm{sc}$ are in better agreement with the experimentally refined structures of orthorhombic and layered \ch{LiMnO2} in the International Crystal Structure Database (ICSD), although the differences between $\mathrm{PBEsol}+(U+V)_\mathrm{sc}$ and $\mathrm{PBEsol}+U_\mathrm{sc}$ are rather small.

\begin{figure*}
    \centering
    \begin{subfigure}{0.48\textwidth}
        \includegraphics[scale=0.58]{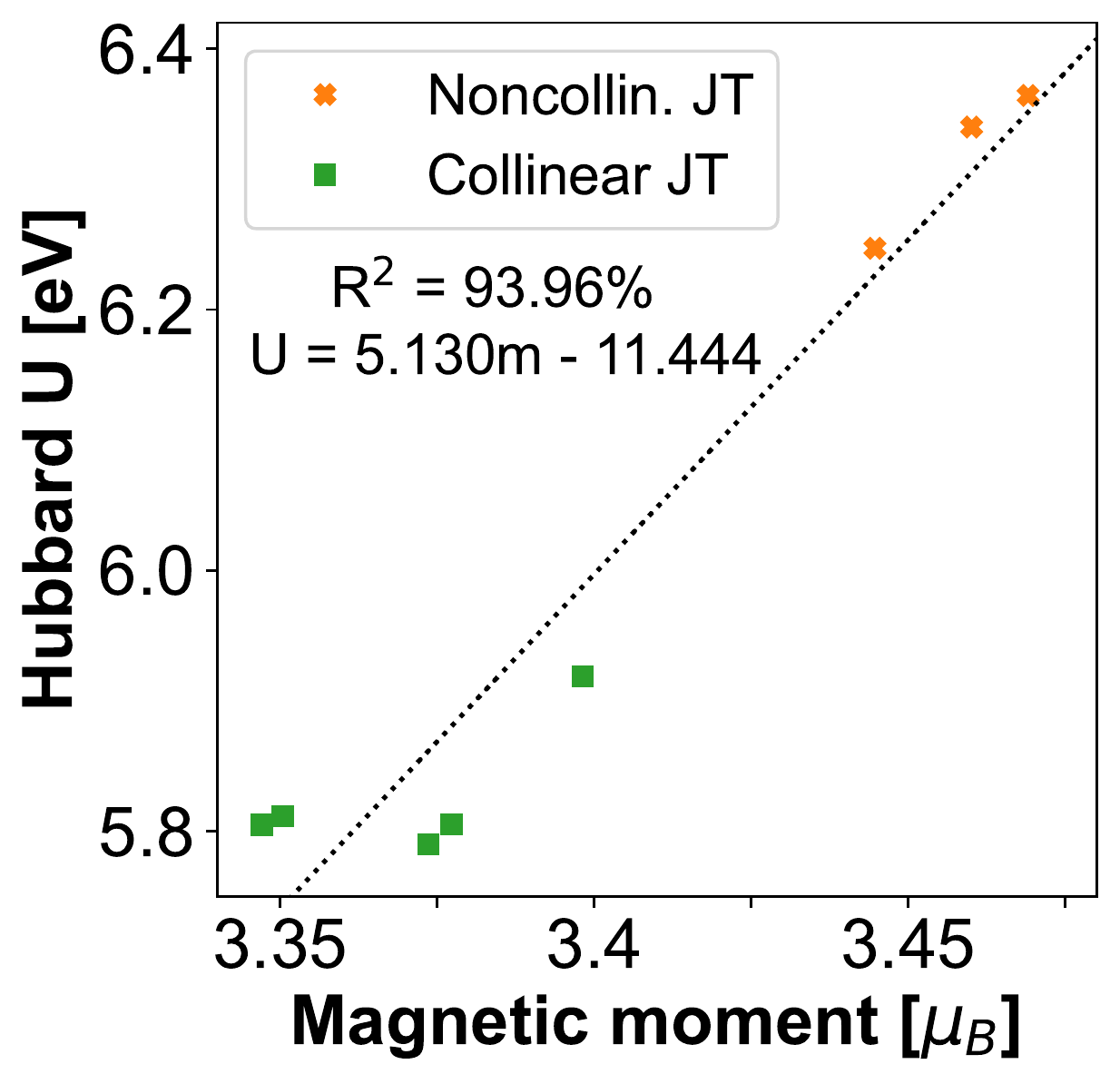}
        \caption{}
        \label{fig:hubb_u_mag}
    \end{subfigure}
    \hfill
    \begin{subfigure}{0.48\textwidth}
        \includegraphics[scale=0.58]{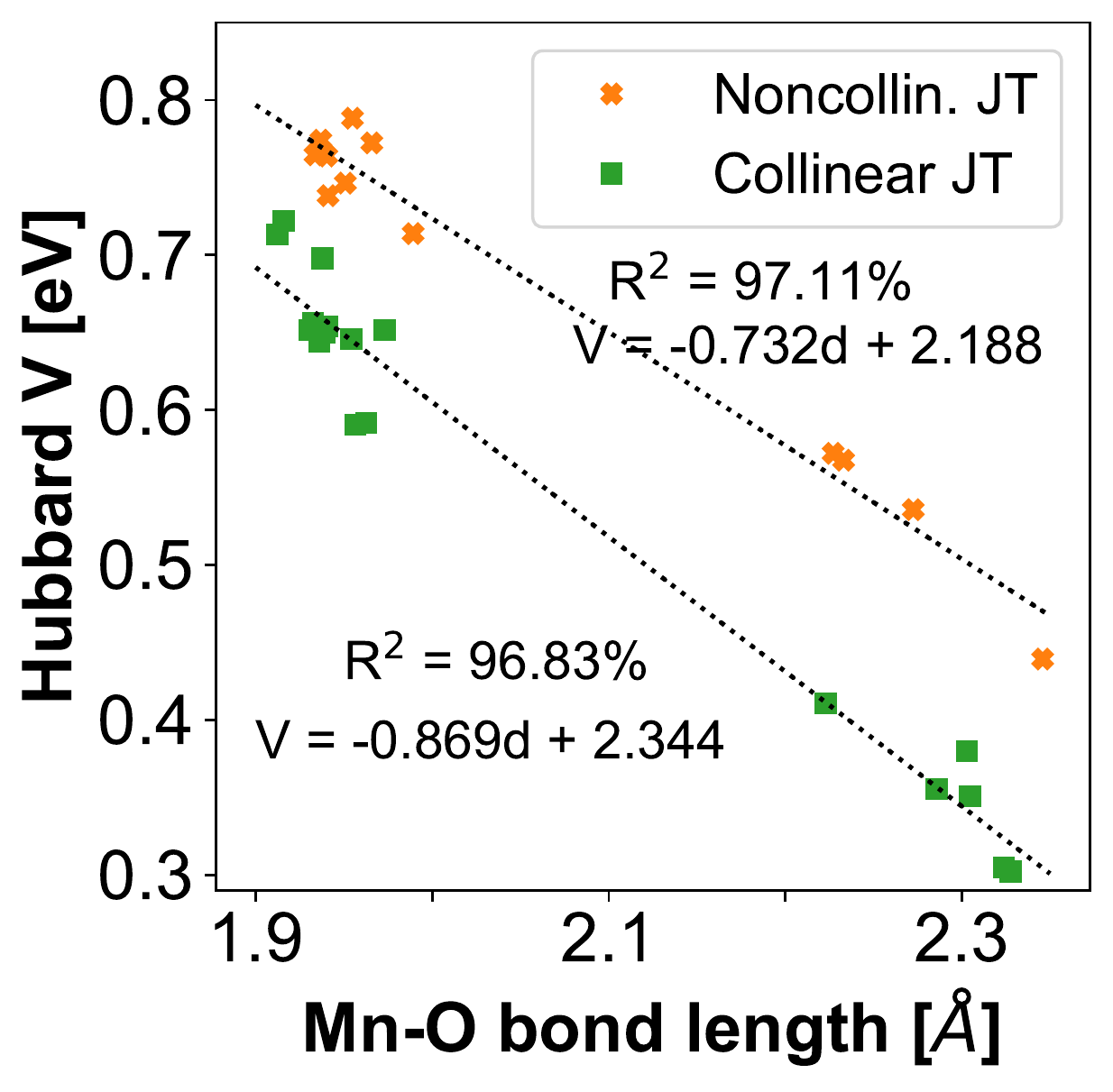}
        \caption{}
        \label{fig:hubb_v_bonds}
    \end{subfigure}

    \caption{Correlating self-consistent HPs from $\mathrm{PBEsol}+(U+V)_\mathrm{sc}$ with the local electronic structure and bonding environments. Plots of a) Hubbard $U$ with Mn magnetic moment and b) $V$ with \ch{Mn}-\ch{O} bond length. All magnetic moments and bond lengths are calculated from the self-consistent HPs. The values of HPs are distinguished by the noncollinear (orange crosses) and collinear (green squares) JT distortions. Lines of best fit and $R^2$ values are shown.}
    
\end{figure*}

The variations in the Hubbard parameters can also be correlated with the local electronic structure and bonding environments. We provide correlation plots between the self-consistent values of $U$ (from $\mathrm{PBEsol}+(U+V)_\mathrm{sc}$) and Mn magnetic moments in Figure \ref{fig:hubb_u_mag}, and the values of $V$ with Mn-O bond length in Figure \ref{fig:hubb_v_bonds}. There is a strongly linear relation between Hubbard $U$ and Mn magnetic moment, with a coefficient of determination $R^2=93.96$\% (Figure \ref{fig:hubb_u_mag}). A similar linear relation is also found for $\mathrm{PBEsol}+U_\mathrm{sc}$  (shown in SI Figure S2 \cite{si}). This strongly linear relation between $U$ and Mn moment indicates that the self-consistent $U$ reflects the degree of electron localization on each Mn site. To demonstrate that the variation in Mn magnetic moments is not an artifact of the distinct values of applied HPs, we compare the Mn magnetic moments calculated using averaged HPs ($U$ = 6 eV and $V$ = 0.6 eV) with the Mn moments calculated using $\mathrm{PBEsol}+(U+V)_\mathrm{sc}$ in SI Figure S6 \cite{si}, which shows that the magnetic moments from both methods are very close (RMSE = 0.008 $\mu_\mathrm{B}$).

The larger values of $U$ specifically arise on \ch{Mn} atoms that are part of 180$\degree$ \ch{Mn}-\ch{O}-\ch{Mn} complexes in which one \ch{Mn}-\ch{O} bond is long ($\sim$ 2.3 \AA) and the other is short ($\sim$ 1.9 \AA). In the following discussion, we will refer to a \ch{Mn} taking part in this type of complex as possessing a ``noncollinear'' JT distortion, while the \ch{Mn} that do not as having a ``collinear'' JT distortion. In Figure \ref{fig:hubb_u_mag}, we see that the Mn atoms with collinear JT (green squares) have significantly smaller values of Hubbard $U$ and magnetic moment ($\mathrm{PBEsol}+(U+V)_\mathrm{sc}$) compared to the Mn with noncollinear JT (orange crosses). In ortho, layered, spinel, and $\epsilon$-\ch{LiMnO2}, the JT axes are all collinearly ordered (structures shown in Figure \ref{fig:all_strucs}), which leads to a smaller $U$ compared to $\gamma$-\ch{LiMnO2}, in which the \ch{Mn} have a noncollinear JT distortion and larger $U$. Given that the elongated Mn-O JT axis reflects a filled $e_g$ orbital, the non-collinear arrangement almost certainly reflects filled-empty orbital ordering, consistent with the higher degree of localization and higher $U$ value.

\begin{figure}[!h]
    \centering
    \includegraphics[scale=0.55]{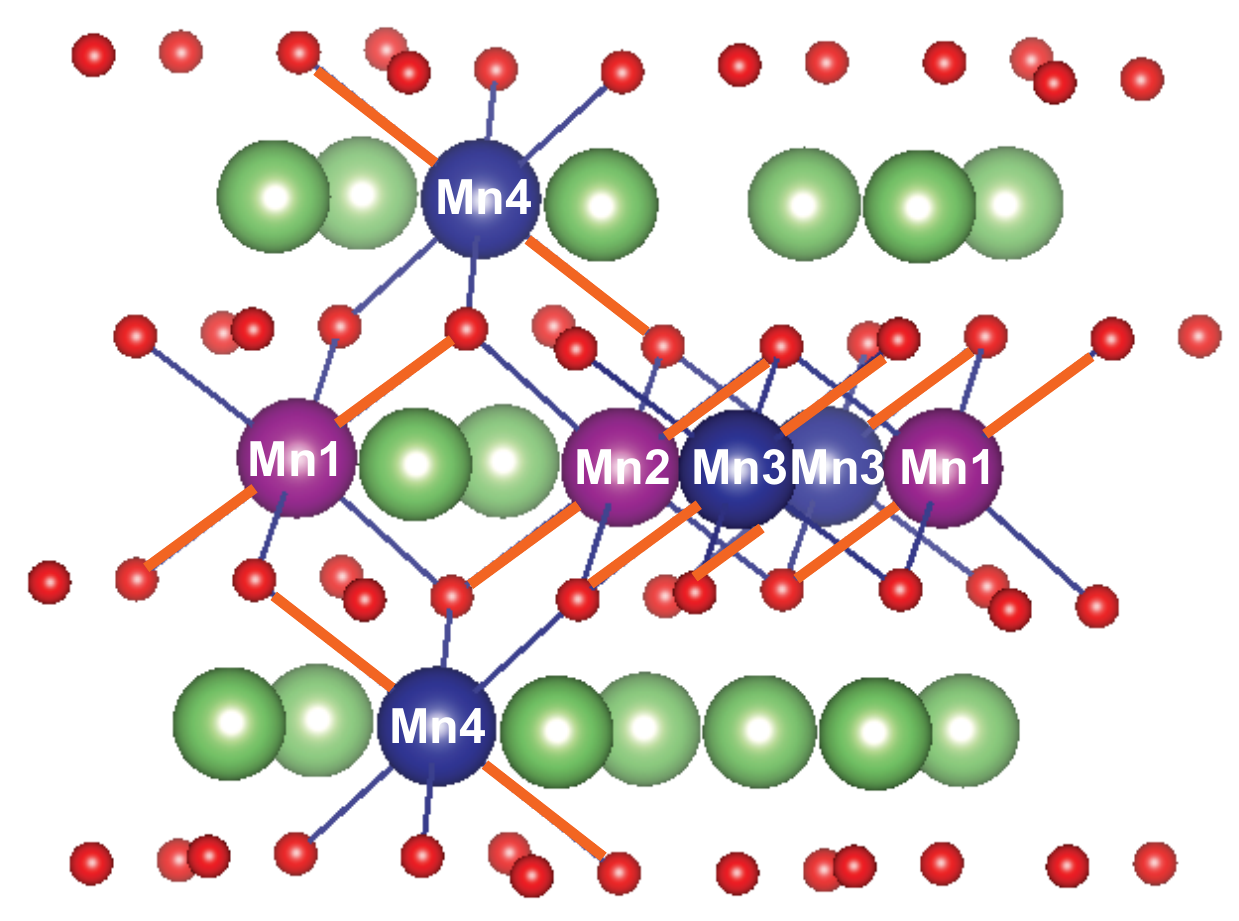}
    \vskip 0.1in
    \renewcommand{\arraystretch}{1.}
    \setlength\tabcolsep{0.26in}
    \begin{tabular}{ccc}
        \toprule
        Mn(\#) & $U$ (eV) & Moment $(\mu_\mathrm{B})$ \\
        \midrule
        Mn1   & 5.92   & $-$3.40   \\
        Mn2   & 5.92   & $-$3.40   \\
        Mn3   & 6.25   & 3.44   \\
        Mn4   & 6.34   & 3.46   \\
        \bottomrule
    \end{tabular}
    \caption{Structure of the disordered layered phase and table of Hubbard $U$ and magnetic moments corresponding to each labeled \ch{Mn}, calculated within $\mathrm{PBEsol}+(U+V)_\mathrm{sc}$. \ch{Mn} atoms are distinguished by their spins (blue for up-spin, and purple for down-spin). Green atoms - Li, red atoms - O.}
    \label{fig:disord_layered_struc}
\end{figure}

In the disordered layered phase, the mixing between \ch{Mn} and \ch{Li} layers leads to distinct \ch{Mn} environments that include both collinear and noncollinear JT distortions in the DFT-relaxed configuration. We display the ionic and magnetic structure of disordered layered \ch{LiMnO2} in Figure \ref{fig:disord_layered_struc}, and tabulate the values of Hubbard $U$ and magnetic moment (within $\mathrm{PBEsol}+(U+V)_\mathrm{sc}$) for each distinct \ch{Mn} site. Mn1 and Mn2 only form 180$\degree$ Mn-O-Li interactions, and thus we can classify these sites as having collinear JT distortions (Figure \ref{fig:disord_layered_struc}). Consequently, Mn1 and Mn2 have relatively small values of $U$ and absolute value of the magnetic moment. Mn3 and Mn4 take part in 180$\degree$ Mn-O-Mn interactions with each other, in which one Mn-O bond is long and the other short - thus both have a noncollinear JT distortion. The values of $U$ on Mn3 and Mn4 are significantly greater than that of Mn1 and Mn2.

We also observe a strongly linear correlation between $V$ and the Mn-O bond length (d$_{\text{Mn-O}}$) (Figure \ref{fig:hubb_v_bonds}). On first glance, there appears to be a large spread in the values of $V$, although the trend of $V$ with d$_{\text{Mn-O}}$ is approximately linear with a R$^2$ of 80\%. The values of $V$ can be partitioned into two groups: one with \ch{Mn} sites that have a noncollinear JT (orange crosses in Figure \ref{fig:hubb_v_bonds}) and the other group consisting of \ch{Mn} with collinear JT distortions (green squares in Figure \ref{fig:hubb_v_bonds}). Within these groups, the linear fits of $V$ as a function of d$_{\text{Mn-O}}$ become remarkably more accurate, with $R^2=97$\%. The values of $V$ associated with \ch{Mn} sites that have a collinear JT are significantly smaller than the $V$ on \ch{Mn} sites with noncollinear JT, as the lines of best fit are separated by $>$ 0.1 eV in the examined range of d$_{\text{Mn-O}}$. Thus the values of Hubbard $U$ and $V$ appear to be correlated to the local \ch{Mn}-$3d$ orbital ordering. The linear trend between $V$ and d$_{\text{Mn-O}}$ can be rationalized as a shorter bond length reflecting stronger binding between the \ch{Mn}-$3d$ and \ch{O}-$2p$ states, and thus leading to a larger hybridizing $V$ value. 

\begin{table*}
    \caption{Magnetic moment of \ch{Mn} and band gap of ortho-\ch{LiMnO2} computed from self-consistent field calculations using various DFT functionals (\texttt{Quantum ESPRESSO}). AFM order is used, unless otherwise stated.}
    \centering
    \begin{adjustbox}{width=\textwidth}
    \begin{tabular}{c c c c c c c}
    \hline 
         Property & PBEsol & PBEsol & PBEsol & PBEsol & PBEsol & HSE06 \\
          & & $+U$ & +$U$+$V$ & +$U$+$V$ (FM) & +$U_\mathrm{Mn,O}$+$V$ & \\
         \hline 
         Mn moment [$\mu_B$] & 3.00 & 3.38 & 3.32 & 3.35 & 3.43 & 3.26 \\
         Band gap [eV] & 0.9 & 1.6 & 2.0 & 1.2 & 2.5 & 3.1 \\
         \hline 
    \end{tabular}
    \end{adjustbox}
    \label{tab:mn_mom_eg}
\end{table*}

\subsection{\label{sec:el_structure}Electronic structure of orthorhombic \ch{LiMnO2}}

We have observed large differences in the \ch{LiMnO2} phase stability trends when different magnetic order and DFT approximations are applied. To rationalize these differences, we analyze the electronic structure of ortho-\ch{LiMnO2}. For these calculations, we fix the atomic coordinates to the primitive cell of the neutron diffraction refined AFM superstructure \cite{greedan-ortho-nd-jssc-1997}. In the following discussion, the HPs imposed on PBEsol are self-consistently calculated at the experimental configuration (shown in SI Table III \cite{si}), which are in good agreement with the values shown in Table \ref{tab:hps}.

\begin{figure}
     \centering
     \includegraphics[scale=0.46]{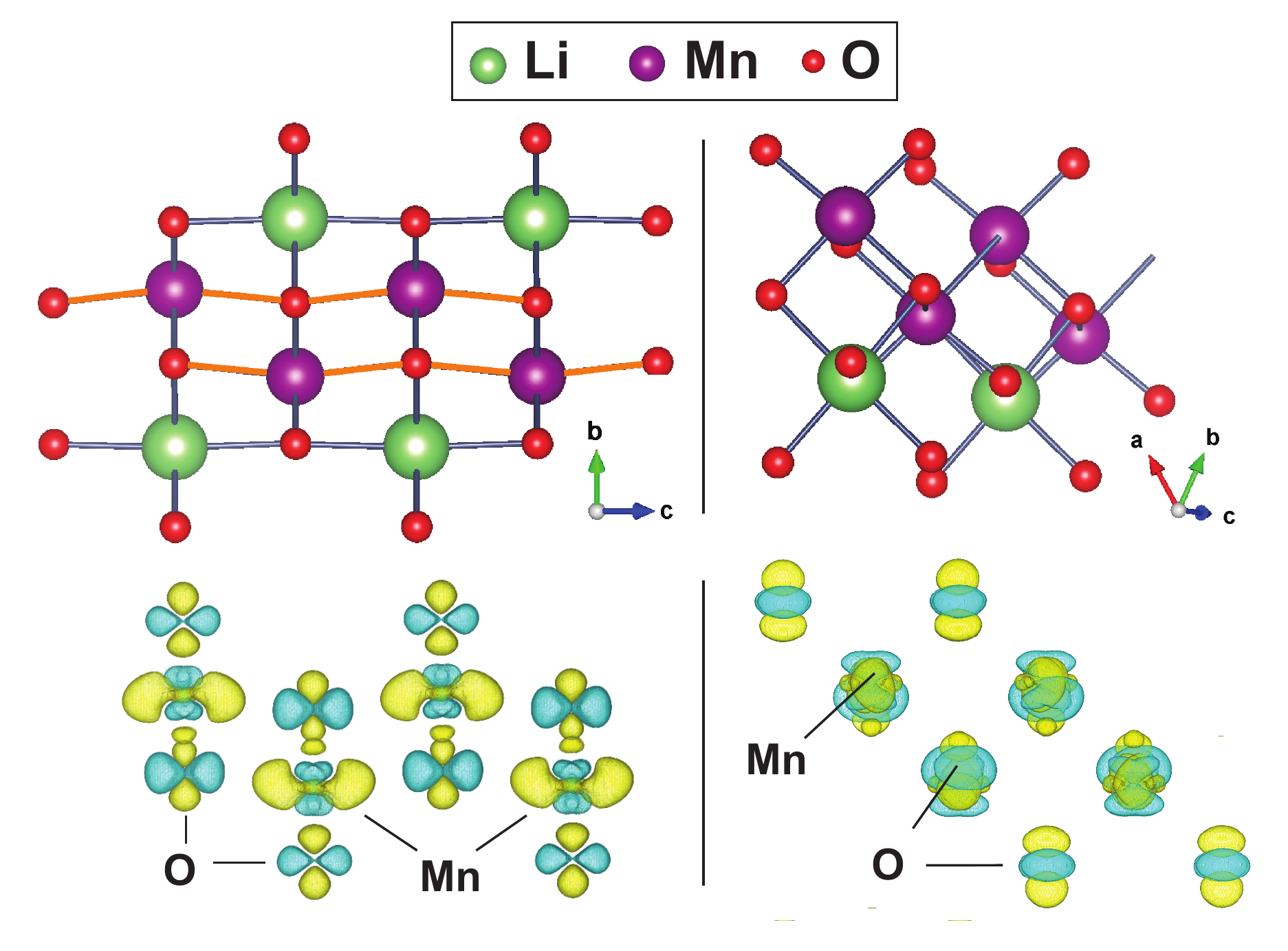}
     \caption{Isosurfaces of the difference in  charge density ($\rho$) between FM and AFM states of ortho-\ch{LiMnO2} ($\rho\mathrm{(AFM)} - \rho\mathrm{(FM)}$) computed within $\mathrm{PBEsol}+U+V$, visualized at two different crystal orientations as shown in the top panels. The yellow (blue) regions indicate areas of increased (decreased) $\rho$ in the AFM structure. The bonds highlighted in orange correspond to the long bond ($\sim$2.3 \AA) in the Jahn-Teller complex.}
     \label{fig:rho_diff}
 \end{figure}

The electron charge density ($\rho$) of ortho-\ch{LiMnO2} is computed in the AFM and FM states using $\mathrm{PBEsol}+U+V$. The magnetic moment and band gap of these calculations are listed in Table \ref{tab:mn_mom_eg}. We compute the difference in $\rho$ between these two calculations ($\rho_\text{AFM}$ - $\rho_\text{FM}$) and plot the isosurfaces in Figure 
\ref{fig:rho_diff}a). In the AFM state, $\rho$ increases predominantly along lobes oriented towards the Mn JT axis in the $\hat{c}$ direction (yellow in Figure \ref{fig:rho_diff}), resembling $e_g$ orbitals. The electron density also decreases (blue in Figure \ref{fig:rho_diff}) around \ch{Mn} along the lobes that are approximately orthogonal to the JT axes, which correspond to $t_{2g}$-like states. Thus, there appears to be a rearrangement of $\rho$ near \ch{Mn}, such that electrons are removed from $t_{2g}$ orbitals and fill the $e_g$ orbitals. The shape of the positive isosurface along the $e_g$ states is more diffuse compared to the more localized $t_{2g}$ orbitals,  indicating a more pronounced covalency between \ch{Mn} and the neighboring \ch{O} along the JT axis. This increased covalency is reflected in the slightly lower \ch{Mn} magnetic moment in the AFM state (3.32 $\mu_B$) compared to the FM case (3.35 $\mu_B$) (Table \ref{tab:mn_mom_eg}). We also observe significant rearrangement of $\rho$ near each \ch{O} site, as $\rho$ is decreased from lobes that are oriented along $\hat{c}$ having a $2p$-orbital shape, and $\rho$ is increased along the $2p$-like lobes oriented along [110]. The AFM order also leads to a significantly larger band gap compared to FM (2.0 and 1.2 eV, listed in Table \ref{tab:mn_mom_eg}). These substantial changes in electronic structure driven by altering the magnetic order should contribute to the large energy differences calculated using different spin configurations.

To understand the differences between the DFT methods considered, we perform SCF calculations of AFM ortho-\ch{LiMnO2} using PBEsol, $\mathrm{PBEsol}+U$, $\mathrm{PBEsol}+U+V$, $\mathrm{PBEsol}+U_\mathrm{Mn,O}+V$ and HSE06, where $U_\mathrm{Mn,O}$ refers to a Hubbard $U$ imposed on both the Mn-$3d$ and O-$2p$ manifolds. The \ch{Mn} magnetic moment and band gap calculated using each method are presented in Table \ref{tab:mn_mom_eg}. 
The band gaps increase from PBEsol to PBEsol$+U$, to $\mathrm{PBEsol}+U+V$, to $\mathrm{PBEsol}+U_\mathrm{Mn,O}+V$. The magnetic moments do not increase in the same order, as $V$ tends to penalize larger on-site moments.
However, all the magnetic moments lie within a reasonable range of each other, from 3 $\mu_\mathrm{B}$ with PBEsol to 3.43 $\mu_\mathrm{B}$ with $\mathrm{PBEsol}+U_\mathrm{Mn,O}+V$.

\begin{figure}
     \centering
     \includegraphics[scale=0.52]{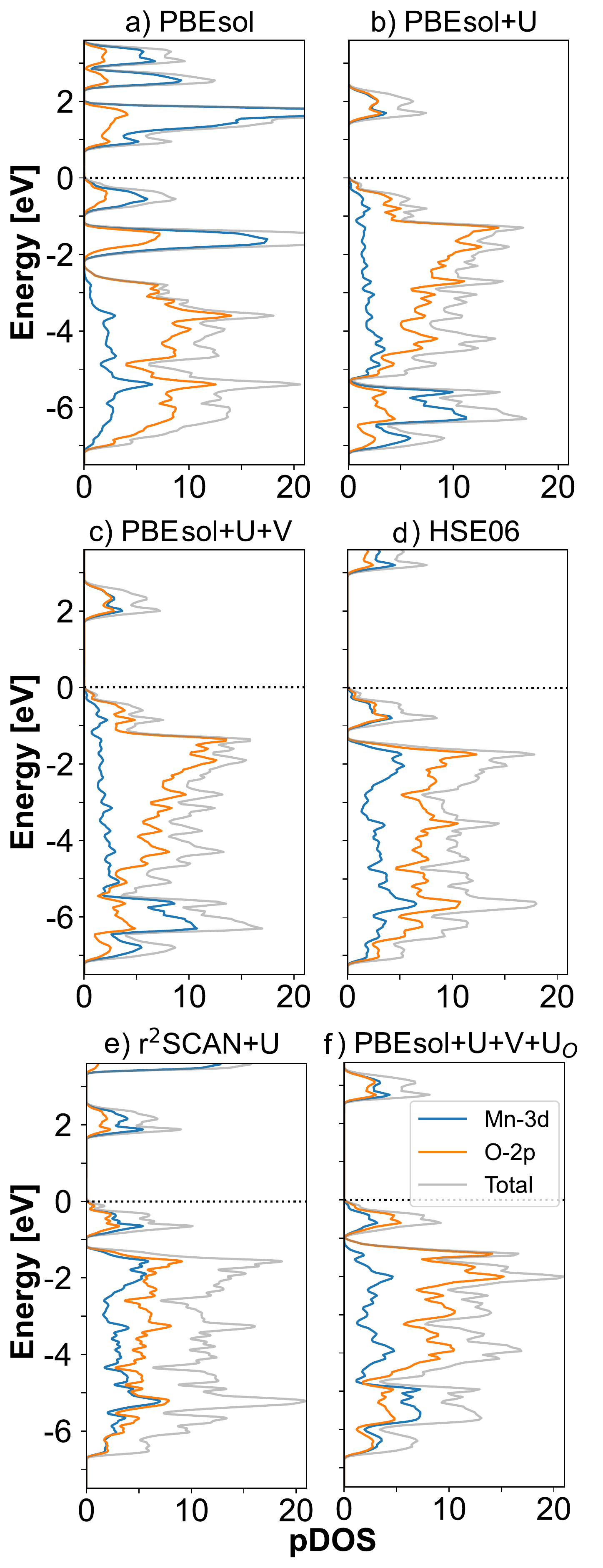}
     \caption{Projected density of states (pDOS) of ortho-\ch{LiMnO2} computed using a) PBEsol, b) $\mathrm{PBEsol}+U$, c) $\mathrm{PBEsol}+U+V$, d) HSE06, e) \rrscan+ $U$ and f) $\mathrm{PBEsol}+U_\mathrm{Mn,O}+V$. Blue: \ch{Mn}-$3d$, orange: \ch{O}-$2p$, and grey: total DOS. Energies are referenced to $E_\mathrm{F}$.}
     \label{fig:pdos_ortho}
 \end{figure}

 The energy levels and orbital character of the electronic states can provide insights about the degree of electronic hybridization and localization. The projected density of states (pDOS) from the scf calculations on ortho-\ch{LiMnO2} are shown in Figure \ref{fig:pdos_ortho}. We examine the energy range from 0 to $-8$ eV relative to the Fermi level ($E_\mathrm{F}$ - which we refer to as the highest occupied energy level) and contains the band manifolds that are mostly \ch{Mn}-$3d$ and \ch{O}-$2p$ in character.

\begin{figure*}
    \centering
    \includegraphics[scale=0.90]{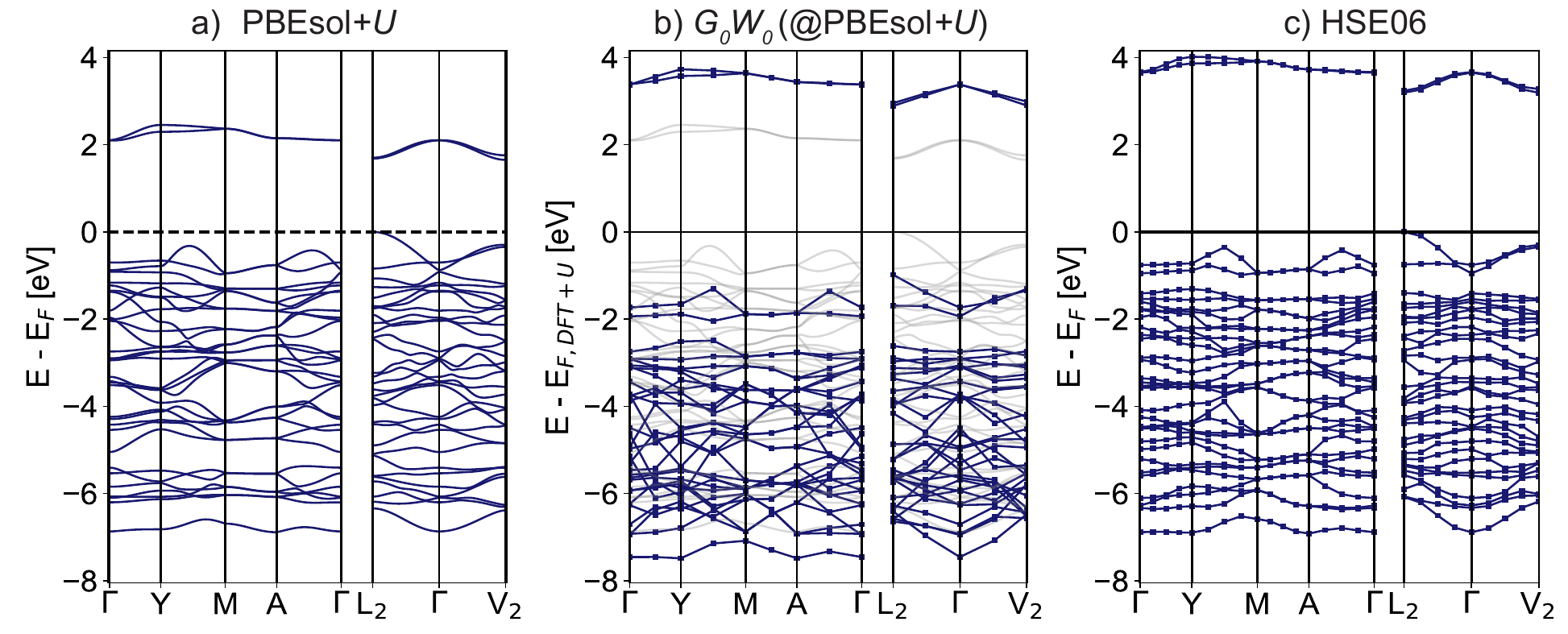}

    \caption{Electronic band structure of the ND-refined AFM ortho-\ch{LiMnO2} structure, computed using a) $\mathrm{PBEsol}+U$, b) $G_0W_0$ starting from the $\mathrm{PBEsol}+U$ wave functions, and c) HSE06. The majority and minority spin channels are degenerate. In b), the band energies are referenced to the $E_\mathrm{F}$ calculated from $\mathrm{PBEsol}+U$ and the $\mathrm{PBEsol}+U$ band structure is shown in light grey to highlight the $G_0W_0$ self-energy corrections. Marked squares in b) and c) indicate the points where the band energies are computed and the lines are linear interpolations between them.}
    \label{fig:bands_ortho}
    
\end{figure*}

 In the calculation using PBEsol, the states in the region between 0 to $-2$ eV are predominantly \ch{Mn}-$3d$ in character (Figure \ref{fig:pdos_ortho}a). The pDOS spectrum shows a sharp peak in the region between $-1$ to $-2$ eV, which indicates that these \ch{Mn}-$3d$-like bands are relatively flat and unhybridized (the PBEsol band structure is shown in SI Figure S4 \cite{si}). When an on-site $U$ is added ($\mathrm{PBEsol}+U$), these top valence states become significantly more \ch{O}-$2p$ in character, while the states at lower energy ($-5$ to $-7$ eV) become more \ch{Mn}-$3d$ in character (Figure \ref{fig:pdos_ortho}b). Thus, the on-site $U$ shifts the \ch{O}-$2p$ states higher in energy relative to \ch{Mn}-$3d$. The peaks in the pDOS at $-1$ to $-2$ eV that are observed in PBEsol (Figure \ref{fig:pdos_ortho}a) are not present in $\mathrm{PBEsol}+U$ (Figure \ref{fig:pdos_ortho}b), which indicates that these bands have greater dispersion when $U$ is applied. Indeed, it can be seen from plots of the band structure that between $-1$ to $-2$ eV, the PBEsol bands (SI Figure S4 \cite{si}) are flatter than $\mathrm{PBEsol}+U$ (Figure \ref{fig:bands_ortho}). The pDOS spectrum computed within $\mathrm{PBEsol}+U+V$ (Figure \ref{fig:pdos_ortho}c) is similar to $\mathrm{PBEsol}+U$, with the main difference being the increase in band gap when adding the inter-site $V$ term (2.0 eV compared to 1.6 eV in $\mathrm{PBEsol}+U$). 

 When HSE06 is employed , the manifold of states in the region between 0 to $-1$ eV becomes separated from the rest of the states that are lower in energy (Figure \ref{fig:pdos_ortho}d).  We assign the states in the 0 to $-1$ eV region to represent the $e_g$ orbitals, while the states lower in energy to be the $t_{2g}$ orbitals, which would be consistent with the crystal field theory of transition metal octahedral complexes. Within this picture, the energy separation between $e_g$ and $t_{2g}$ states represents the crystal field splitting energy ($\Delta E_\text{cf}$). Within HSE06, the $t_{2g}$ states (from $-5$ to $-7$ eV) have slightly more \ch{O}-$2p$ character than \ch{Mn}-$3d$, and the difference between the two is nearly uniform along the entire energy range. The more uniform distribution of \ch{Mn}-$3d$ and \ch{O}-$2p$-like states in HSE06 suggests an increased \ch{Mn}-\ch{O} hybridization and covalency.
   
 The \rrscan+ $U$ pDOS (Figure \ref{fig:pdos_ortho}e) shows a band gap (1.7 eV) that is substantially smaller than the HSE06 band gap (3.1 eV). However, the occupied manifold of \rrscan+ $U$ is in striking agreement with HSE06; in particular, both \rrscan+ $U$ and HSE06 display energetically separated $e_g$ and $t_{2g}$ states (i.e. a nonzero $\Delta E_\mathrm{cf}$) -- a feature that is not present neither in $\mathrm{PBEsol}+U$ nor in $\mathrm{PBEsol}+U+V$. Interestingly, within PBEsol, a nonvanishing $\Delta E_\mathrm{cf}$ can be recovered by applying an additional Hubbard $U$ on the O-$2p$ states (see Figure \ref{fig:pdos_ortho}f -- $\mathrm{PBEsol}+U_\mathrm{Mn,O}+V$). Thus, the manifestation of the crystal field splitting appears to be intimately related to the electronic localization on O-$2p$ states.

 Since we have assessed a large array of DFT methods, it is important to evaluate the accuracy of each approximation. Given a lack of experimental data on the electronic structure of ortho-\ch{LiMnO2}, we perform a $G_0W_0$ calculation as a reference point for a realistic description of the electronic structure. The band structure of ortho-\ch{LiMnO2} is computed using $\mathrm{PBEsol}+U$, $G_0W_0$, and HSE06, which are all shown in Figure \ref{fig:bands_ortho}. The bands of the majority and minority spin channels are degenerate. To display the differences between the electronic states calculated within $G_0W_0$ and its $\mathrm{PBEsol}+U$ starting point (i.e. the self-energy corrections), we reference the band energies of these two calculations to $E_\mathrm{F}$ of the $\mathrm{PBEsol}+U$ calculation in Figures \ref{fig:bands_ortho}a and \ref{fig:bands_ortho}b.

Within $G_0W_0$ (Figure \ref{fig:bands_ortho}b), we observe a large renormalization of the $\mathrm{PBEsol}+U$ band energies (Figure \ref{fig:bands_ortho}a), with self-energy corrections on the order of 1 eV in magnitude, though the values of the self-energies are highly dependent on the band and wave vector. The two empty $e_g$ conduction bands are rigidly shifted upward in energy by $\sim$ 1.2 eV and the valence bands are shifted down by up to $>$ 1 eV compared to $\mathrm{PBEsol}+U$ (Figure \ref{fig:bands_ortho}a), leading to a significantly larger band gap (3.8 eV) compared to $\mathrm{PBEsol}+U$ (1.6 eV). 
 There are more band crossings in the $t_{2g}$ manifold within $G_0W_0$, which highlights the larger dispersion of these bands compared to $\mathrm{PBEsol}+U$. The crystal field splitting $\Delta E_{cf}$ is $\sim$ 0.3 eV between the $e_g$ and $t_{2g}$ bands. In $\mathrm{PBEsol}+U$, $\Delta E_\mathrm{cf}$ vanishes as an $e_g$ and $t_{2g}$ band are degenerate at $\Gamma$ (Figure \ref{fig:bands_ortho}a), which is consistent with the $\mathrm{PBEsol}+ U$ pDOS of Figure \ref{fig:pdos_ortho}a. The values of $\Delta E_\mathrm{cf}$ and the $e_g$ band widths in $G_0W_0$ are close to those in the HSE06 band structure (Figure \ref{fig:bands_ortho}c) and \rrscan+ $U$ DOS (Figure \ref{fig:pdos_ortho}f), which validates the electronic structure of the valence states predicted by these DFT functionals.

\section{\label{sec:discussion}Discussion}

In this study, we have computed the total energy of the \ch{LiMnO2} polymorphs using several DFT methods and evaluated harmonic phonon contributions to the free energy at finite temperature. Obtaining accurate phase stability trends is not trivial in this system, as precise description of the AFM order and interplay between electronic localization/hybridization are required. These effects are inadequately treated within more conventional DFT functionals (e.g. $\mathrm{PBEsol}+U$, \rrscan, and \rrscan+ $U$, where values of $U$ are empirically tuned), which lead to the spurious prediction that $\gamma$-\ch{LiMnO2} is the ground state. Instead, DFT schemes such as HSE06, $\mathrm{PBEsol}+U_\mathrm{sc}(+V_\mathrm{sc})$, and \rrscan+ $U_\mathrm{sc}$ are needed to recover ortho-\ch{LiMnO2} as the ground state. Furthermore, these functionals are necessary to predict a disordered layered ordering to be higher in energy than the layered phase, which represents an appreciable anti-site defect formation energy in layered.

We find that an important aspect of the energy difference between all structures arises from the local variation of self-consistently determined Hubbard parameters, especially the on-site $U$, which is why schemes with averaged Hubbard parameters cannot reproduce the correct phase stability. Specifically, the self-consistent $U$ and $V$ are smallest in the ortho, layered, spinel, and $\epsilon$ phases (Table \ref{tab:hps}). These phases all exhibit a collinear ordering of the JT distortions, in which all of the JT axes (and hence the $e_g$ orbitals) are oriented in parallel. Capturing these local variations in electronic interactions in DFT is key for obtaining accurate energetics in this system, and likely would be important for modeling other TM oxides rich in Mn$^{3+}$ or other JT-active ions (e.g. Ni$^{3+}$ and Cu$^{2+}$). HSE06 can also reasonably capture these local differences in the electron-electron interactions despite not explicitly applying these Hubbard corrections, as it predicts orthorhombic to be the \ch{LiMnO2} ground state. Range-separated hybrid functionals such as HSE06 more rigorously treat the screened electronic exchange interactions as they incorporate a fraction (0.25 in HSE06) of exact Fock exchange to the short-range exchange energy, which can provide an adequate correction to the self-interaction error (SIE) \cite{hse06-2006, heyd-screen-hybrid-2003}. This indicates that a major source of the errors encountered in applying $\mathrm{PBEsol}+U$ and \rrscan($+U$) is the poor description of electronic exchange, which can be remedied by calculating the HP self-consistently. 

The differences in formalism of the hybrid-GGA and GGA with self-consistent HPs lead to a different phase stability trend between the $\gamma$ and ortho phases: within HSE06, $\gamma$ is only marginally higher in energy ($\sim$ 1 meV/atom, Figure \ref{fig:dft_uv_hse_gs_ens}) than ortho-\ch{LiMnO2}, while $\mathrm{PBEsol}+U_\mathrm{sc}(+V_\mathrm{sc})$ predict a much larger energy difference ($>$ 20 meV/atom). It is difficult to conclude which method is most accurate for this quantity, as there is no known experimental measurement. However, we note that within HSE06, we apply values of the fraction of exact exchange ($\alpha$ = 0.25) and range-separation parameter ($\omega$ = 0.2) that may not be the optimal values for each \ch{LiMnO2} phase. Methods have been proposed to non-empirically and systematically tune the parameters within hybrid-GGA functionals to more realistically describe the electronic screening and/or minimize the SIE within a given structure \cite{wing-tune-hybrid-2021, otrsh_layered_2024, chen_dielec_hybrid_2018, sagredo_hybrid_2025} -- in the same spirit as the calculation of HPs from linear response \cite{cococcioni-ldau-prb-2004, cococcioni-ggauv-2010}. Thus, it may be insightful in future studies to evaluate the LiMnO2 phase stability using a non-empirically tuned hybrid-GGA, though this would come at larger computational cost.

The improved treatment of electronic exchange within HSE06 enables it to more accurately predict the band structure of ortho-\ch{LiMnO2}, as it shows reasonable agreement with the many-body $G_0W_0$ method, especially in the band gap and crystal field splitting ($\Delta E_{cf}$) (Figure \ref{fig:bands_ortho}). These quantities are more poorly described within $\mathrm{PBEsol}+U_\mathrm{sc}(+V_\mathrm{sc})$, which we have attributed to an insufficient correction of SIE on O-$2p$ states, which can be alleviated by applying an additional Hubbard $U$ correction on O-$2p$ states ($U_\mathrm{O}$), \rrscan + $U$, or HSE06. The values of $U_\mathrm{O}$ that we calculate from DFPT are $\sim$ 9 eV (SI Table S3 \cite{si}), which are larger than the $U$ values of Mn ($\sim$ 6 eV), but in good agreement with previous self-consistent calculations of $U_\mathrm{O}$ within GGA functionals \cite{moore-ggauj-prm-2024, wayne-dftu-eg-2021}. The $G_0W_0$ and HSE06 methods predict large band gaps (3.8 and 3.1 eV, respectively) that are significantly higher than previously reported calculations \cite{mishra-lmo-stability-prb-1998, Singh1994, hoang_limno2_hse_defects, galakov_limno2_lsda_pes, banerjee-limno2-dmft-2023}, and indicate that these phases are strongly insulating.

The electronic structure of these phases are heavily linked to the magnetic order, as the AFM order is shown to increase the covalency of \ch{Mn}-\ch{O} bonds, which likely contributes to the significant lowering of energy in each AFM structure relative to the FM state (Figure \ref{fig:afm_fm_en_diff}). This enhancement in the hybridization between the Mn-$3d$ and O-$2p$ states is correlated with an improved prediction of phase stability (Figure \ref{fig:ggau_r2scan_gs_ens}b). Correlation between AFM interactions and Mn-O covalency is consistent with the theory of AFM "semicovalent exchange", which was initially developed by Goodenough and Loeb to rationalize the antiferromagnetism of various spinel and perovskite phases \cite{goodenough-afm-spinels-1954, goodenough-afm-perov-1958, goodneough-perovskites-1955}. Within this theory, a local AFM interaction would lead to the formation of semicovalent bonds between a TM and anion \cite{goodenough-afm-spinels-1954, goodenough-afm-perov-1958, anderson-exchange-1963, kanamori-exch-sym-1959}. The large energy difference between FM and AFM states in the ortho and $\epsilon$ structures ($\Delta E_{AFM-FM} \approx$ $-80$ to $-100$ meV/Mn, shown in Figure \ref{fig:afm_fm_en_diff}) suggests that the semicovalent exchange is particularly strong in these phases and is an important factor that helps stabilize them relative to the other \ch{LiMnO2} phases. The ion configurations of ortho and $\epsilon$-\ch{LiMnO2} (Figure \ref{fig:all_strucs}) contain $90\degree$ and $180\degree$ Mn-O-Mn channels for AFM interactions, whereas layered and spinel only have 90$\degree$ Mn-O-Mn channels. This structural difference may be why AFM order can have a stronger stabilizing effect on ortho and $\epsilon$ compared to the other experimentally-observed layered and spinel phases.

Despite the disparity of phase stability trends observed across different DFT functionals, the energy differences between the experimentally known ortho, layered, and spinel \ch{LiMnO2} phases are surprisingly consistent across each method, which suggests a similarity in the electronic structure of these phases. The $\mathrm{PBEsol}+U(+V)$, \rrscan $(+U)$, and HSE06 functionals all predict that the order of DFT total energy from low to high energy is ortho $\rightarrow$ layered $\rightarrow$ spinel, when all structures are in their respective AFM ground state (Figures \ref{fig:afm_gs_ens} and \ref{fig:dft_uv_hse_gs_ens}). The energy relative to ortho is $5-15$ meV/atom for spinel and $3-8$ meV/atom for layered, depending on the functional used. The vibrational free energy of these phases between $T$ = 0 to 1000 K are also very similar (Figure \ref{fig:tot_f_energy}). These observations theoretically confirm that these phases are very thermodynamically competitive across a wide range of temperatures. This thermodynamic competition elucidates why many experimental procedures to synthesize ortho \ch{LiMnO2} can lead to spinel and/or layered impurity phases \cite{jang-al-limno2-ssi-2000, reimers-lt-limno2-jecs-1993, gummow-limno2-synth-jecs-1994, gummow-lt-ortho-mrb-1993}. Since the energy differences at the \ch{LiMnO2} composition are small, minor off-stoichiometry in Li or Mn content could possibly lead to changes in the phase stability. If this is the case, then the presence of impurity phases may be a result of particles containing off-stoichiometry or following a somewhat different synthesis path. Indeed, it has been previously shown from experiment and computation that upon delithiating layered or ortho \ch{LiMnO2}, there is a driving force to form the spinel phase (see SI Figure S9) \cite{mishra-lmo-stability-prb-1998, armstrong-lay-spin-chem-mat-2004, 10.1149/1.1838013, reed_layered_spinel_2001, mishra-lmo-stability-prb-1998, paulsen_lmo_pd_air, reed_layered_spinel_2001, jain_commentary_2013}.

Our calculations reveal that the previously unreported $\epsilon$-\ch{LiMnO2} is a potential low-energy phase, since it is predicted to have an energy comparable to all of the experimentally-known phases across all levels of DFT assessed. Specifically, the $\epsilon$ structure is consistently predicted to be only $\sim$ 2 meV/atom higher in energy than ortho and lower in energy than layered and spinel. These results may be surprising, as the $\epsilon$ cation ordering has never been reported. The low energy of the $\epsilon$ phase can be rationalized from its similarities in its structural and electronic properties with ortho, layered, and spinel. Namely, we have shown that $\epsilon$ has a collinear ordering of JT axes, comparable Mn magnetic moment, and similar self-consistent Hubbard parameters to these phases.

\begin{figure*}
    \centering
    \includegraphics[scale=0.4]{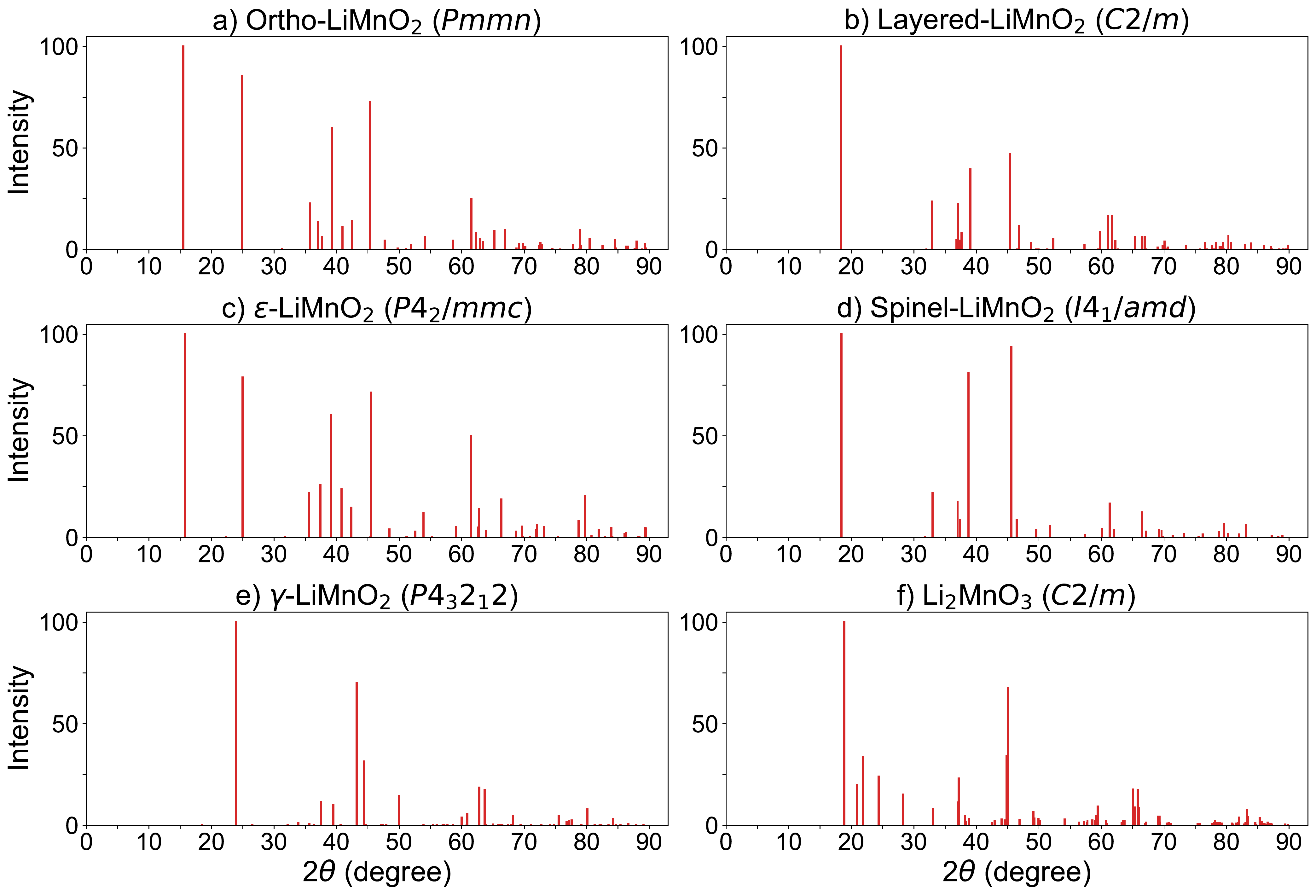}

    \caption{Simulated XRD (Cu K-$\alpha$ radiation) patterns of \ch{LiMnO2} and \ch{Li2MnO3} phases. Each structure is relaxed within HSE06 and the identified space group of the final structure is shown in parentheses.}
    \label{fig:xrd}
    
\end{figure*}

To further evaluate the structural similarity of the ortho and $\epsilon$ phases, we compute their XRD patterns, which are shown in Figures \ref{fig:xrd}a and \ref{fig:xrd}c, respectively. Both $\epsilon$ and ortho-\ch{LiMnO2} share the same high intensity peaks at 2$\theta$ = 15$\degree$, 25$\degree$, 39$\degree$, 45$\degree$, and 62$\degree$. The key differences in these patterns are in the disappearance of one of two peaks at 37$\degree$ for $\epsilon$, and different relative intensities among the peaks between 35-45$\degree$. While a low energy above the ground state is no guarantee for synthesizability \cite{wenhao_metastable_2016}, its similar diffraction pattern compared to ortho may have led $\epsilon$ to be missed in previously reported \ch{LiMnO2} samples.

The $\gamma$ phase is predicted to be the ground state within the PBEsol and r$^2$SCAN functionals, with or without averaged Hubbard corrections (Figures \ref{fig:fm_gs_ens} and \ref{fig:afm_gs_ens}). Since the $\gamma$ ordering has never been reported in the \ch{LiMnO2} composition from experiment, we would expect it to be higher in energy than the experimentally-found phases. We simulate the XRD pattern of $\gamma$ and display it in Figure \ref{fig:xrd}e, which clearly does not resemble the XRD patterns of ortho, layered, or spinel (Figures \ref{fig:xrd}a, \ref{fig:xrd}b, and \ref{fig:xrd}d, respectively). Thus, $\gamma$ would almost certainly not be mistaken for either of these phases, unlike the case of $\epsilon$. The $\gamma$ structure corresponds to the cation ordering with the lowest electrostatic energy among \ch{Li$M$O2} rock salt structures \cite{wu-limo2-philmag-1998}. Hence, the spurious \ch{LiMnO2} ground state prediction is an indication that these methods inadequately capture the specific energy contributions that stabilize ortho-\ch{LiMnO2}, and instead overly stabilize electrostatically-favorable configurations. Our results show that the phase stability and electronic structure of \ch{LiMnO2} are more strongly influenced by the subtle balance between electron localization and hybridization, instead of electrostatics. The HSE06, $\mathrm{PBEsol}+U_\mathrm{sc}(+V_\mathrm{sc})$, and \rrscan + $U_\mathrm{sc}$ methods appear to provide a more faithful treatment of these aspects, enabling them to correctly predict ortho-\ch{LiMnO2} as the ground state.

\begin{table*}
    \caption{Phonon entropy [$
    k_B$/f.u.] of the \ch{LiMnO2} phases as a function of temperature.}
    \centering
    \begin{adjustbox}{width=\textwidth}
    \begin{tabular}{c c c c c c c c }
    \hline 
         Temp. [K] & Ortho & Layered & Spinel & $\epsilon$ & $\gamma$ & Disord. Layered & SQS \\
         \hline 
         300 & 8.26 & 8.25 & 8.29 & 8.26 & 7.79 & 7.94 & 7.97 \\
         600 & 15.45 & 15.45 & 15.49 & 15.44 & 14.97 & 15.14 & 15.19 \\
         900 & 20.08 & 20.09 & 20.13 & 20.08 & 19.61 & 19.78 & 19.82 \\
         \hline
    \end{tabular}
    \end{adjustbox}
    \label{tab:ph_s}
\end{table*}

Our phonon calculations reveal that $\gamma$ becomes more unstable relative to ortho as temperature is increased (Figure \ref{fig:tot_f_energy}), due to its lower vibrational entropy. Disordered layered is also destabilized by its lower phonon entropy, but to a smaller extent than $\gamma$. Since $\gamma$ and disordered layered are the only phases considered that contain noncollinear arrangements of JT distortions, we speculate that the vibrational properties are closely linked to the degree and type of cation and orbital ordering. To further investigate this correlation, we generate a disordered \ch{LiMnO2} special quasi-random structure (SQS) (methods described in the SI Section VI \cite{si}) \cite{zunger_sqs, stochastic_sqs_2013} and compare the structural and vibrational properties to the aforementioned \ch{LiMnO2} phases. Upon structural relaxation within $\mathrm{PBEsol}+U$, SQS-\ch{LiMnO2} contains a noncollinear arrangement of JT distortions (structure shown in SI Figure S7 \cite{si}). We compute the phonon entropy ($S_{\text{ph}}$) of SQS-\ch{LiMnO2} as a function of temperature and compare it to the other phases, as presented in Table \ref{tab:ph_s} (also plotted in SI Figure S8 \cite{si}). At 300 K, SQS-\ch{LiMnO2} has a very comparable $S_{\text{ph}}$ compared to disordered layered and $\gamma$. These three phases have a noticeably lower $S_{\text{ph}}$ (by $0.3 - 0.5$ $k_B$/f.u.) compared to ortho, layered, spinel, and $\epsilon$-\ch{LiMnO2}, which have very similar $S_{\text{ph}}$ at 300 K (within 0.05 $k_B$/f.u. of each other). These trends are consistent across the temperature range examined. These findings suggest that only certain cation configurations can favor the formation of collinear $e_g$ orbital arrangements, which in turn contributes to increasing the $S_{\text{ph}}$.

Since orthorhombic \ch{LiMnO2} is the ground state and also exhibits relatively large $S_{ph}$, the collinear JT ordering appears to provide a unique source of both electronic and vibrational stability. We speculate that the relatively softer phonon modes in these structures can arise from increased phonon anharmonicity \cite{samara_ktao3_anharm} and/or electron-phonon (el-ph) coupling \cite{zhang-coo2-doping-2004, zhang-cuprate-epc-2007}, effects that are associated with cooperative JT distortions \cite{kanamori-jt-1960, radin_jt_order_2020, millis-jt-lamno3-1996, pavarini-lamno3-jt-2010}. Although our employed frozen phonon method does not explicitly account for anharmonicity and el-ph coupling, it may capture small contributions from these effects.

Our findings also indicate that the degree of electron localization and hybridization can be affected by the ordering of JT distortions, as shown by our identification of large variations in self-consistent HPs. The self-consistent Hubbard $U$ in $\gamma$ is significantly larger than that of the experimentally-known phases (Table \ref{tab:hps}). The values of $U$ are shown to linearly correlate with the Mn magnetic moment (Figure \ref{fig:hubb_u_mag}). This correlation can be physically intuitive, since when the electron density becomes more confined on Mn-$3d$ orbitals, the strength of the on-site Coulomb interactions ($U$) between these electrons can naturally increase as well, leading to a larger energy penalty for localization. Greater electron localization on Mn implies a smaller degree of hybridization between Mn and O neighbors, suggesting that the nature of the Mn-O bonding within $\gamma$ is more ionic, and the bonding within the experimentally-known phases is more covalent. Our calculated electronic pDOS of ortho-\ch{LiMnO2} indeed suggests that the Mn-O bonding in this phase is highly covalent in nature, as there is a strong hybridization of Mn-$3d$ and O-$2p$ states within the valence $e_g$ and $t_{2g}$ manifolds near the Fermi level, especially within \rrscan + $U$ and HSE06 (Figure \ref{fig:pdos_ortho}). The disordered layered structure also contains Mn with noncollinear JT distortions and larger values of $U$, which reflects how the formation of anti-site defects in layered \ch{LiMnO2} can decrease the covalency of some Mn-O bonds. Only the more precise treatments of electronic exchange and correlation within HSE06 and $\mathrm{PBEsol}+U_\mathrm{sc}(+V_\mathrm{sc})$ can reasonably capture the increase in energy arising from reduced Mn-O covalency, which enables the prediction of significant anti-site defect formation energies.

\begin{figure}
    \centering
    \includegraphics[width=0.95\linewidth]{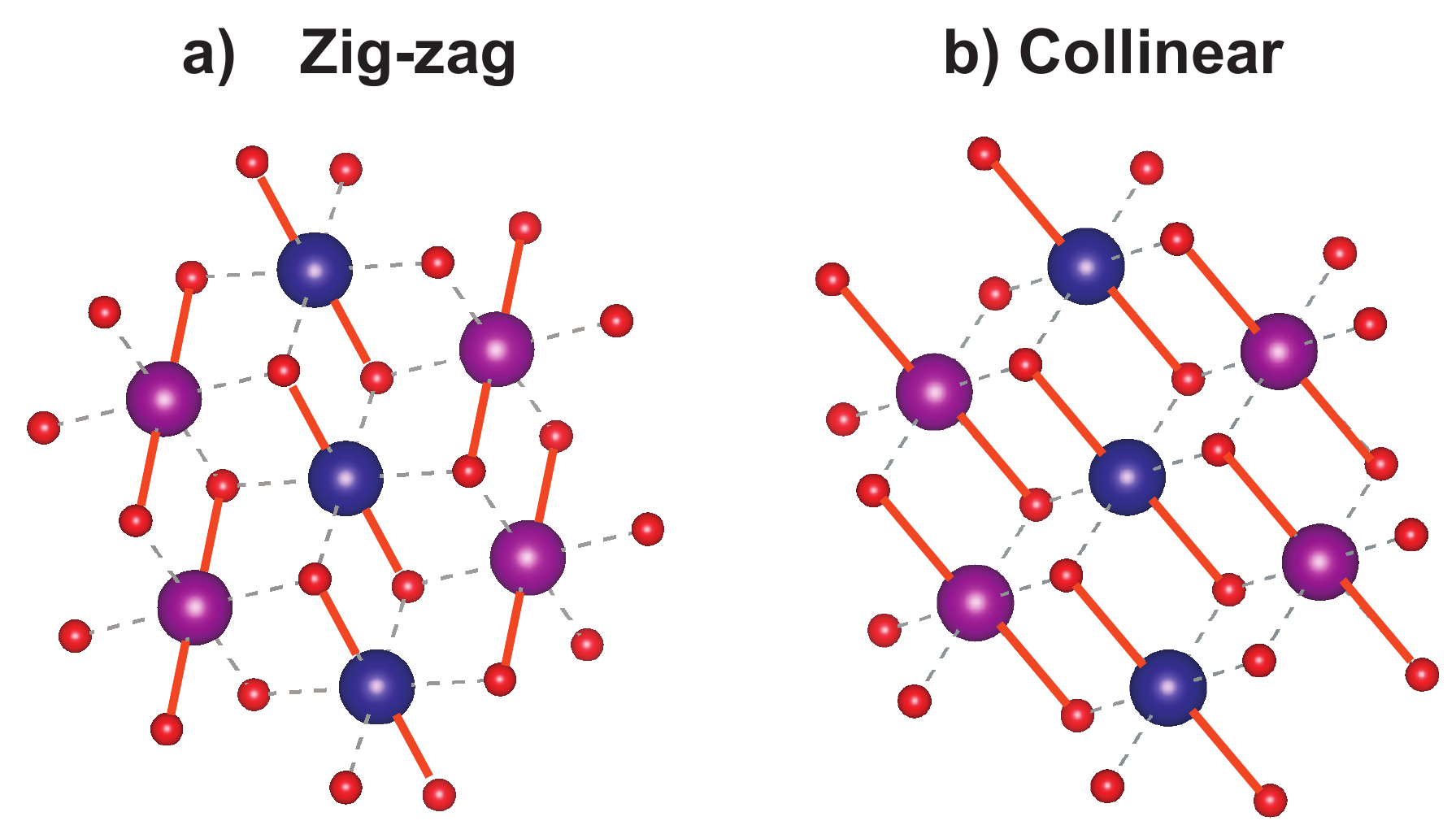}
    \caption{Structures of AFM layered \ch{LiMnO2} with a) zig-zag (P2$_1$/c) and b) collinear (C2/m) ordering of JT axes. Only one Mn layer is shown for clarity. The JT axes are highlighted in orange. Blue - Mn (up spin), purple - Mn (down spin), and red - O.}
    \label{fig:layered_jt_orderings}
\end{figure}

To clarify the impact of orbital ordering on the electronic properties, we calculate the properties of layered \ch{LiMnO2} with a zig-zag arrangement of JT distortions instead of a collinear arrangement -- we will refer to this structure as zz-layered \ch{LiMnO2} (P2$_1$/c space group), which is shown in Figure \ref{fig:layered_jt_orderings}. The zig-zag ordering of JT axes has been previously studied in layered \ch{LiMnO2} and \ch{LiNiO2} \cite{10.1021/acs.chemmater.7b03080,genreith-jt-lno-2023}. We calculate the HPs, Mn magnetic moment, and JT bond ratio of zz-layered within $\mathrm{PBEsol}+(U+V)_\mathrm{sc}$, and compare them to layered \ch{LiMnO2} in Table \ref{tab:layered_zz_comp}. Indeed, we find that zz-layered has a self-consistent Hubbard $U$ that is $\sim$ 170 meV larger than layered \ch{LiMnO2}, as well as a larger magnetic moment. In fact, if we use the line of best fit shown in Figure \ref{fig:hubb_u_mag} and the computed Mn moment of zz-layered, we can predict a Hubbard $U$ of 6.05 eV, which differs from the self-consistent $U$ only by 0.09 eV. Furthermore, the JT bond ratio of zz-layered is much lower than that of the collinear JT structure, and comparable to $\gamma$-\ch{LiMnO2} (Table \ref{tab:hps}). These effects stemming from the collinear JT ordering are thermodynamically favorable, as the computed total energy of the collinear JT ordering is $64-140$ meV/Mn lower than the zig-zag JT arrangement, depending on the DFT method used (SI Table S4 \cite{si}). These results provide further evidence that collinear $e_g$ orbital ordering facilitates greater Mn-O covalency and enhances the JT distortion, which results in lowering the total energy.

\begin{table*}[]
\caption{Comparing the properties of layered \ch{LiMnO2} with collinear and zig-zag orderings of JT axes -- self-consistent Hubbard parameters, Mn moments, and JT bond ratio ($\frac{\text{long}}{\text{avg. short}}$) within $\mathrm{PBEsol}+(U+V)_\mathrm{sc}$.}
    \label{tab:layered_zz_comp}
\renewcommand{\arraystretch}{1.0}
    \setlength\tabcolsep{0.29in}
    \centering
    \begin{tabular}{c c c c c}
    \hline
         Jahn-Teller & Hubbard & Hubbard & Mn & JT bond  \\
          Ordering & $U$ [eV] & $V$ [eV]  & Moment [$\mu_B$] & ratio \\
         \hline
        Collinear (C2/m) & 5.79 & 0.65, 0.30 & 3.38 & 1.203 \\
        Zig-zag (P2$_1$/c) & 5.96 & 0.68, 0.60, 0.37 & 3.41 & 1.147 \\
        \hline
    \end{tabular}
    
\end{table*}

Our results suggest that the orbital ordering is particularly influential to the phase stability of \ch{LiMnO2}. All structures with a collinear JT ordering exhibit increased covalency of Mn-O bonding, magnitude of JT distortion (Table \ref{tab:hps}), energy differences between the AFM and FM states (up to 100 meV/Mn in ortho and $\epsilon$, shown in Figure \ref{fig:afm_fm_en_diff}), and vibrational entropy (Table \ref{tab:ph_s}). Thus, the JT ordering appears critically linked to several aspects of the electronic and vibrational sources of stability. The \ch{LiMnO2} structures refined from experiment all exhibit the collinear JT ordering \cite{greedan-ortho-nd-jssc-1997, akimoto_limno2_drx, hoppe_ortho_1975, 10.1149/1.1838013,armstrong-layered-limno2-nature-1996, capitaine-layered-limno2-ssi-1996, david-li2mn2o4-nd-jssc-1987}, which further indicates that this cooperative JT effect is indeed very thermodynamically favorable in \ch{LiMnO2}.
  
  We remark that the stability of the collinear JT ordering and its link to other contributions of the phase stability are not ubiquitous to all TM oxides with JT-active ions. In the low-temperature orthorhombic \ch{LaMnO3} perovskite phase, experimental structural refinements have shown that the JT arrangement is noncollinear \cite{lamno3-struc-1997, eriksson_lamno3_refine}. In previous studies of layered \ch{LiNiO2}, DFT ($\mathrm{GGA}+U$) calculations have shown that the noncollinear zig-zag JT ordering is the ground state orbital arrangement \cite{radin-jt-2017, genreith-jt-lno-2023}. Radin et al. further showed from $\mathrm{GGA}+U$ calculations that the lowering of energy resulting from the JT distortion in layered \ch{LiNiO2} ($\sim -120$ meV/Ni) is much smaller in magnitude than in layered \ch{LiMnO2} ($\sim -350$ meV/Mn) \cite{radin-jt-2017}. Furthermore, the stabilizing effect of AFM order in layered \ch{LiNiO2} is predicted to be minimal ($\Delta E_{AFM-FM} \approx$ $-3$ meV/Ni) \cite{radin-jt-2017} compared to layered \ch{LiMnO2} ($\Delta E_{AFM-FM} \approx$ $-40$ to $-60$ meV/Mn), which may be largely due to the higher spin state of the Mn$^{3+}$ electron configuration compared to Ni$^{3+}$. 
  
  These discrepancies between layered \ch{LiMnO2} and \ch{LiNiO2} suggest that the contributions to the phase stability of these systems are significantly different. This is indeed reflected in the distinct ground states of these systems --- for \ch{LiNiO2} it is the layered structure \cite{genreith-jt-lno-2023, wu-limo2-philmag-1998}, while for \ch{LiMnO2} it is the orthorhombic phase.
   The orthorhombic phase is unique, as it is the ground state only for \ch{LiMnO2}, while the layered phase is the ground state for many other \ch{Li\textit{M}O2} compositions (\textit{M} = Cr, Co, V, etc) \cite{wu-limo2-philmag-1998}. These observations suggest that the \ch{LiMnO2} phase stability is likely dictated by contributions that are unique to this system, and specifically, the Mn$^{3+}$ ion. Our analysis indicates that these contributions are the strong AFM exchange interactions, collinear JT ordering, increased Mn-O covalency, softening of phonon modes, and the coupling between these factors. The strength of JT distortions and AFM interactions are much greater in \ch{LiMnO2} than in \ch{LiNiO2}, so we suspect that the strength of these contributions may be unique to \ch{LiMnO2}, and indeed helps underpin the unusual phase behavior of this system.

\section{Conclusion}
We have revealed the challenges of modeling the phase stability and electronic structure of the \ch{LiMnO2} polymorphs, and identified avenues for obtaining more accurate results in this system. The degree of electron localization on Mn-$3d$ states is found to vary significantly by phase, and is closely correlated to the self-consistently calculated Hubbard $U$ on Mn-$3d$ states. These values of $U$ can vary by $\sim$ 0.6 eV within PBEsol, and are closely linked to the local JT ordering. More conventional DFT approaches (e.g. GGA and meta-GGA with or without averaged Hubbard corrections) inadequately treat the energy differences arising from subtle changes in electron localization and Mn-O bonding, leading to inaccurate phase stability predictions. Instead, more precise treatments of electronic exchange and correlation (and, especially, screened exchange) within functionals such as HSE06, $\mathrm{PBEsol}+U_\mathrm{sc}(+V_\mathrm{sc})$, and \rrscan + $U_\mathrm{sc}$ are required to recover more accurate phase stability -- namely the prediction of orthorhombic \ch{LiMnO2} to be the ground state. We expect these DFT methods to be similarly well-equipped to capture the energetics of other TM oxides that are rich in Mn$^{3+}$ or other JT-active ions (e.g. Ni$^{3+}$ and Cu$^{2+}$).

The ordering of the cations, Mn spins, and $e_g$ orbitals all significantly influence the degree of electron localization and hybridization, which dictate the electronic and vibrational free energy contributions. The collinear arrangement of JT distortions, which is present in ortho, layered, spinel, and $\epsilon$-\ch{LiMnO2}, is shown to simultaneously increase the Mn-O covalency, strength of AFM semicovalent exchange, and vibrational entropy compared to structures with noncollinear JT orderings. The subtle interplay between the electron localization, magnetism, JT distortion, and total energy within the \ch{LiMnO2} phases makes this system a potentially useful test case for benchmarking novel density functional approximations.

An extensive re-examination of the electronic structure of ortho-\ch{LiMnO2} reveals the importance of sufficiently correcting the SIE on O-$2p$ states, in addition to Mn-$3d$ states, in order to obtain accurate features of the band structure, such as the crystal field splitting and band gap. These quantities are well-captured within HSE06, as the predicted band structure is in reasonable agreement with $G_0W_0$.

\section{Acknowledgements}
We would like to thank Prof. Steven Louie, Jack McArthur, and other organizing members of the BerkeleyGW workshop for insightful discussion and guidance on $GW$ calculations. We also thank Tucker Holstun and Dr. Annalena Genreith-Schriever  for valuable discussion about the \ch{LiMnO2} structures and Jahn-Teller distortions, respectively. This work was funded by the U.S. Department of Energy, Office of Science, Office of Basic Energy Sciences, Materials Sciences and Engineering Division under Contract No. DE-AC02-05-CH11231 (Materials Project program KC23MP). This research used computational resources of the National Renewable Energy Laboratory (NREL) and the National Energy Research Scientific Computing Center (NERSC).

\bibliography{refs.bib}

\clearpage

\setcounter{page}{1}
\setcounter{section}{0}
\setcounter{table}{0}
\setcounter{figure}{0}

\renewcommand{\thepage}{S\arabic{page}}
\renewcommand{\thesection}{S\arabic{section}}
\renewcommand{\theequation}{S\arabic{equation}}
\renewcommand{\thetable}{S\arabic{table}}
\renewcommand{\thefigure}{S\arabic{figure}}

\onecolumngrid
\section*{Supplementary Information}

\section{I: Structural relaxations and total energy calculations}

The details of the structural relaxations and total energy calculations are listed in Table \ref{table:relax_details}. These differences in DFT calculation parameters within VASP are motivated by the PBE-based ``\texttt{MPRelaxSet}'' and \rrscan-based ``\texttt{MPScanRelaxSet}'' input sets in pymatgen \cite{ong_pymatgen_2013}. The larger cutoffs used within QE calculations compared to VASP are due to the different pseudopotentials used. Within QE calculations, the pseudopotentials are obtained from the Standard Solid-State Pseudopotentials (SSSP) Efficiency PBEsol library (version 1.3.0) and the calculation parameters chosen are consistent with previous studies \cite{prandini2018precision}.

\begin{table}
\centering
\begin{tabular}{|c|c|c|c|c|c|c|}
    \hline
     Functional & Code & Pseudo- & Energy & Energy & Forces & Hubbard \\
      & & potentials & cutoff [eV] & conv. [eV] & conv. [eV/\AA] & $U$ [eV] \\
     \hline
     PBEsol+$U$ & VASP & PBE-PAW & 520 & 1e-6 & 1e-2 & 3.9 \\ \hline
     r$^2$SCAN & VASP & PBE\_52 & 680 & 1e-6 & 2e-2 & \\ \hline
     r$^2$SCAN+$U$ & VASP & PBE\_52 & 680 & 1e-5 & 2e-2 & 1.8 \\ \hline
     HSE06 & VASP & PBE\_52 & 520 & 1e-5 & 3e-2 & \\ \hline
     PBEsol+$U$+$V$ & QE & SSSP v1.3.0 & 952 & 1.36e-8  & 2e-2 & Table I \\
      & & PBEsol efficiency & (70 Ry) & (1e-9 Ry) & (4e-4 a.u.) & (main text) \\
     \hline
\end{tabular}

\caption{Calculation details of structural relaxations performed within VASP and QE. The Hubbard $U$, if applied, is imposed on the Mn-$3d$ manifold. }
\label{table:relax_details}
\end{table}

\begin{figure}[!t]
    \centering
    \begin{subfigure}{0.95\textwidth}
        \centering
        \includegraphics[scale=0.6]{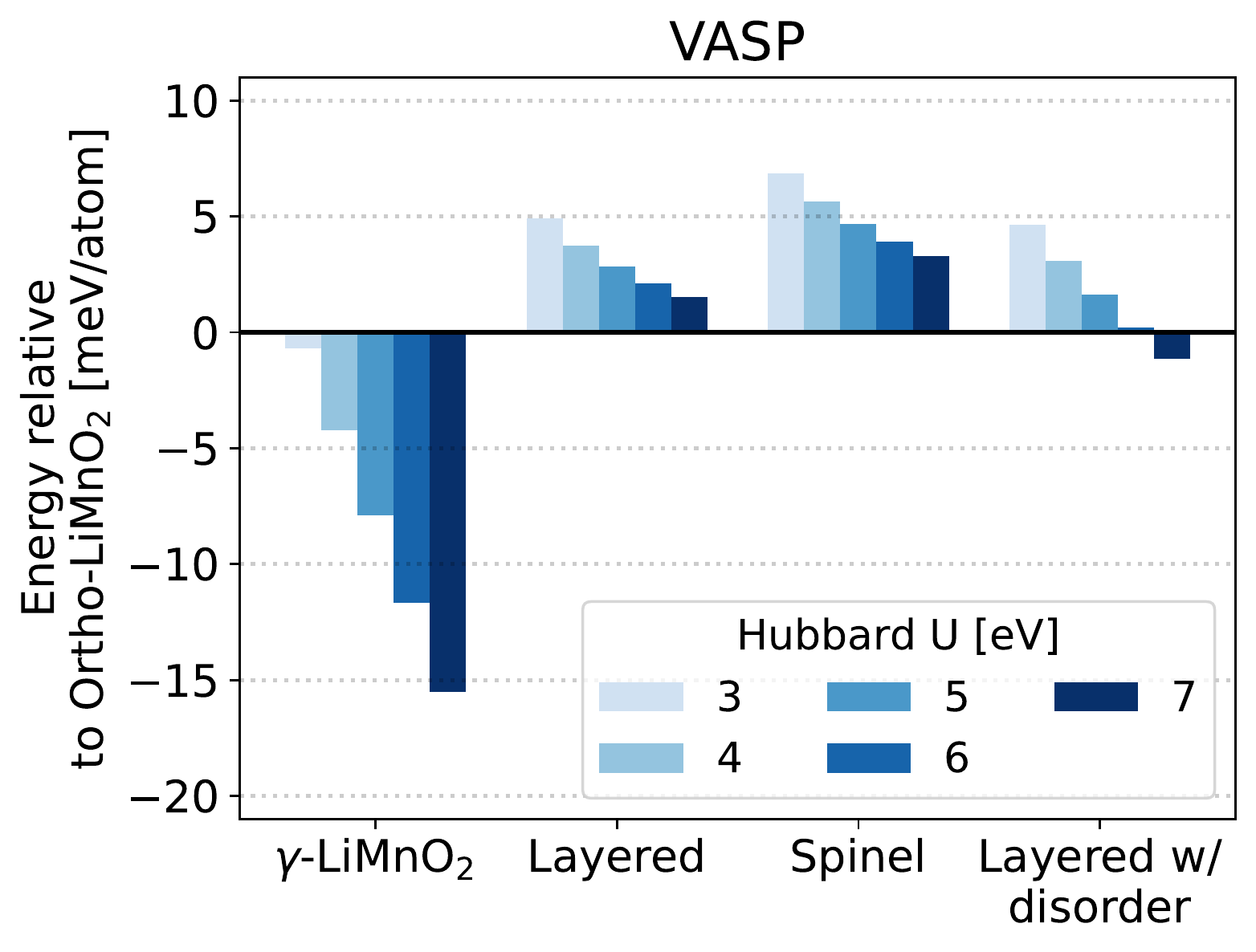}
        \caption{}
        \label{fig: vasp_ens_u}
        
    \end{subfigure}

    \begin{subfigure}{0.95\textwidth}
        \centering
        \includegraphics[scale=0.6]{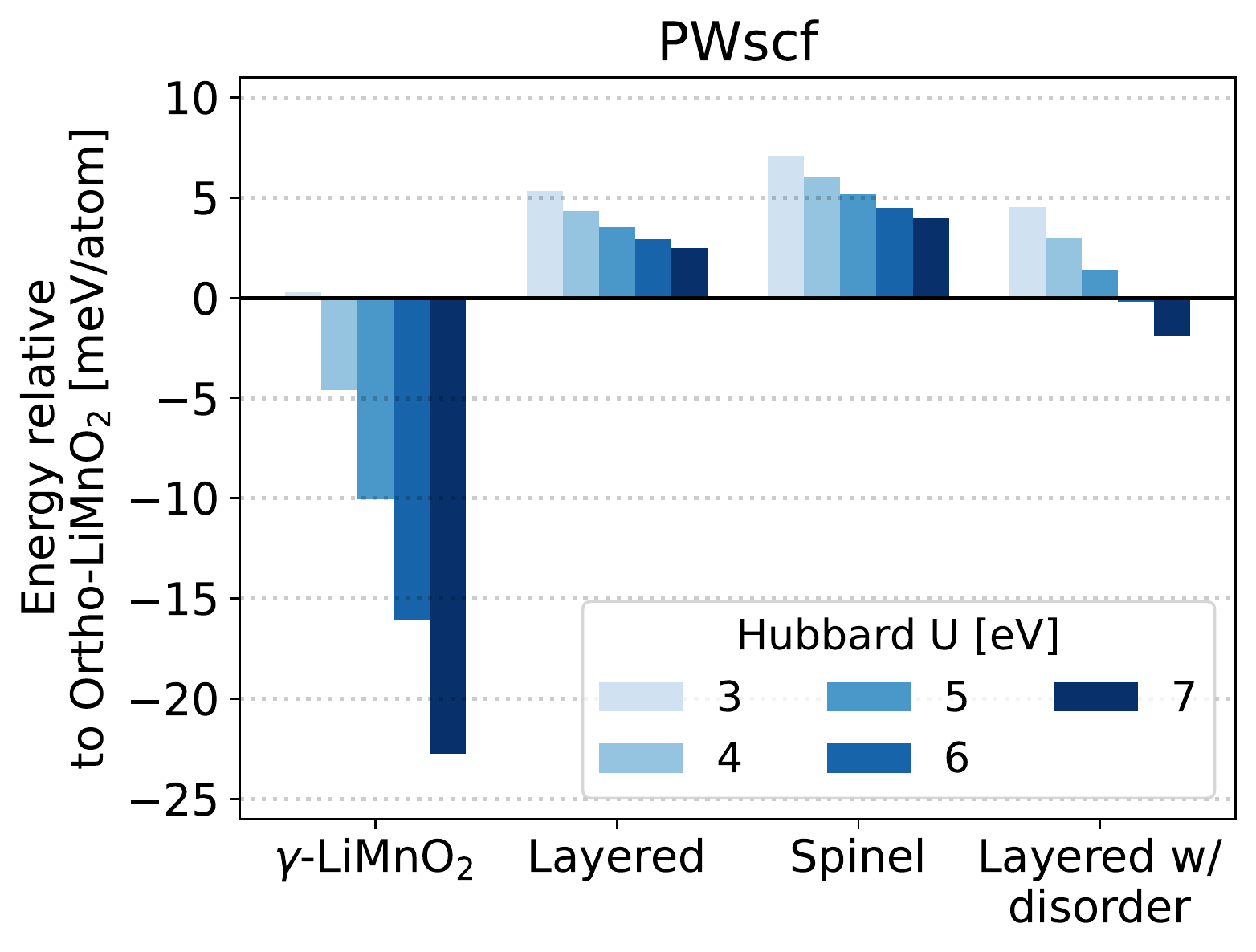}
        \caption{}
        \label{fig: vasp_ens_u}
        
    \end{subfigure}
    \caption{Comparison of computed energies of the \ch{LiMnO2} phases relative to orthorhombic \ch{LiMnO2} within PBEsol+$U$ across a) VASP and b) QE (PWscf). Hubbard $U$ is imposed on Mn-$3d$ manifolds. Each calculation is a SCF calculation.}
    \label{fig:enter-label}
\end{figure}

\section{II: Self-consistent calculation of Hubbard parameters}

We use density functional perturbation theory (DFPT) as implemented in the HP package of QE to calculate the Hubbard $U$ and $V$ parameters within PBEsol from first-principles \cite{qe-2009, qe-2017, timrov-hp-2022}. Lowdin orthogonalized atomic orbitals are used as the Hubbard projectors \cite{timrov2020pulay}. The $\bm{q}$-point grids are chosen to be commensurate with a supercell containing lattice parameters that are $>$ 8 \AA{} and the response function is converged to within 10$^{-6}$ atomic units. Each DFPT calculation requires the ground-state electron density and wavefunctions, which are computed from static self-consistent field (SCF) calculations within $\mathrm{PBEsol}+U+V$. In the first iteration, we use $U$ = 6 eV and $V$ = 0.6 eV, and for the subsequent iterations we use the $U$ and $V$ determined from DFPT in the previous iteration. The SCF calculations are converged to within 10$^{-12}$ Ry (1.36 $\times$ 10$^{-11}$ eV) in the energy. The calculated values of Hubbard $U$ within $\mathrm{PBEsol}+U_{sc}$ are systematically larger than in $\mathrm{PBEsol}+(U+V)_{sc}$. This leads to a systematic increase in the Mn magnetic moments calculated from $\mathrm{PBEsol}+U_{sc}$ as well. In Figure \ref{fig:hubb_u_mag_pbesol_comp}, we plot the self-consistent Hubbard $U$ and resulting Mn moments within each phase, using $\mathrm{PBEsol}+U_{sc}$ and $\mathrm{PBEsol}+(U+V)_{sc}$. Although both methods yield a strongly linear relation between self-consistent $U$ and Mn moment, the lines of best fit differ between each method.

\begin{figure}
    \centering
    \includegraphics[width=0.5\linewidth]{250326_hubb_u_mag_u_uv.pdf}
    \caption{Plots of self-consistent Hubbard $U$ and Mn magnetic moments within each \ch{LiMnO2} phase, calculated using $\mathrm{PBEsol}+U_{sc}$ and $\mathrm{PBEsol}+(U+V)_{sc}$. Lines of best fit and R$^2$ values are also displayed.}
    \label{fig:hubb_u_mag_pbesol_comp}
\end{figure}

For the self-consistent calculations of Hubbard $U$ within \rrscan, we use the linear response constrained DFT (LR-cDFT) approach in real-space \cite{cococcioni-ldau-prb-2004} instead of DFPT, since the current version of \texttt{QE} (version 7.4) does not support DFPT calculations in \rrscan, as well as other codes to our knowledge. Furthermore, the current version of \texttt{QE} does not support variable-cell relaxations within \rrscan, so our procedure is to iteratively calculate the self-consistent $U$ using \texttt{QE}, then perform structural relaxations using VASP with the self-consistent $U$. We expect there to be slight incompatibilities with the implementations of Hubbard $U$ between the two codes, which is the main reason why we leave the results in the SI. In the LR-cDFT calculations, we create isolated perturbations of the Hubbard manifolds using supercells (of 128 atoms and with each lattice parameter $>$ 8 $\AA$) and evaluate the response matrices ($\chi$) of the Hubbard occupations using finite differences. The SCF calculations of the unperturbed and perturbed states are converged to within 1E-11 Ry. The current version of \texttt{QE} only supports norm-conserving pseudopotentials for \rrscan calculations, so we use optimized norm-conserving Vanderbilt (ONCV) PBEsol pseudopotentials selected from the PseudoDojo table \cite{pseudodojo-2018} (version 0.4.1), since \rrscan norm-conserving pseudopotentials have not yet been generated. We note that these ONCV pseudopotentials are different from the PAW and ultra-soft pseudopotentials (selected from SSSP) used for the PBEsol DFPT calculations, which will contribute to differences in the self-consistent HPs \cite{timrov-hp-dfpt-prb-2018, timrov-hp-2022}. Thus, we recompute the Hubbard $U$ within PBEsol+$U_\mathrm{sc}$ using these ONCV pseudopotentials and compare them to the \rrscan + $U_\mathrm{sc}$ values in Table \ref{tab:hp_r2scan}. The plane-wave basis set cutoff for the LR-cDFT calculations within \rrscan (using ONCV pseudopotentials) is 55 Ry (748 eV) and 220 Ry (2993 eV) for the charge density. The final calculated total energy of each \ch{LiMnO2} phase within \rrscan+ $U_{sc}$ used plane-wave cutoffs of 70 Ry (952 eV) and 280 Ry (3810 eV) for the charge density. The Hubbard $U$, Mn magnetic moments, and JT bond ratios computed within \rrscan and PBEsol using ONCV pseudopotentials are shown in Table \ref{tab:hp_r2scan}. The Hubbard $U$ calculated within \rrscan are systematically smaller than PBEsol by $\sim 1$ eV.

We show the \ch{LiMnO2} phase stability computed within \rrscan+ $U_{sc}$ and compare it to the other DFT methods in Figure \ref{fig:phase_r2scan_gga_sc}. The $\gamma$ phase is predicted to be $\sim$ 15 meV/atom higher than ortho, and the disordered layered phase is $\sim$ 10 meV/atom higher than layered, which qualitatively agree with the HSE06 and $\mathrm{PBEsol}+U_{sc}(+V_{sc})$ results. The energy differences among the ortho, layered, and spinel phases are in excellent agreement among all functionals assessed.

\begin{table*}
\caption{Self-consistent Hubbard $U$, Mn magnetic moments, and JT bond ratios of the \ch{LiMnO2} phases within \rrscan and PBEsol, calculated using ONCV pseudopotentials.}
    \centering
    \begin{adjustbox}{width=\textwidth}
    \begin{tabular}{c |c | c | c | c | c | c | c }
    \hline 
         Method & Property & Ortho & Layered & Spinel & $\gamma$ & Disord. layered & $\epsilon$ \\
         \hline 
         \rrscan & $U$ [eV] & 4.48 & 4.46 & 4.47 & 4.62 & 4.45, 4.60, 4.60 & 4.50 \\
         + $U$ & Mag. moment [$\mu_B$] & 3.23 & 3.26 & 3.26 & 3.35 & 3.27, 3.34, 3.34 & 3.23 \\
          & JT ratio & 1.158 & 1.168 & 1.165 & 1.129 & 1.128, 1.174, 1.120 & 1.160 \\
         \hline
         PBEsol & $U$ [eV] & 5.44 & 5.44 & 5.45 & 5.85 & 5.51, 5.77, 5.83 & 5.44 \\
         + $U$ & Mag. moment [$\mu_B$] & 3.34 & 3.36 & 3.36 & 3.48 & 3.38, 3.46, 3.47 & 3.33 \\
         \hline
    \end{tabular}
    \end{adjustbox}
    \label{tab:hp_r2scan}
\end{table*}

\begin{figure}
    \centering
    \includegraphics[width=0.75\linewidth]{250403_dft_ens_hse_dftuv_sc_edge_r2scan_conv.pdf}
    \caption{DFT energy of each \ch{LiMnO2} phase (with AFM configuration) relative to the orthorhombic phase, within several DFT approximations.}
    \label{fig:phase_r2scan_gga_sc}
\end{figure}

\section{III: Phonon calculations}
Phonon calculations are performed within the harmonic approximation using the frozen phonon method, using VASP and Phonopy \cite{kresse_efficient_1996, kresse1993, kresse1994, kresse-paw-1998, togo_phonopy_2015}. Each \ch{LiMnO2} phase is relaxed within its respective AFM ground-state using tighter tolerances - 10$^{-8}$ eV in the energy and 10$^{-4}$ eV/$\AA$ in the forces. The supercell sizes are chosen such that each lattice parameter is $\geq$ 9 \AA. For each supercell snapshot, we perform SCF calculations that are converged to within 10$^{-9}$ eV in the energy using the blocked-Davidson minimization algorithm \cite{kresse_efficient_1996}. The relaxation and SCF calculations are performed within $\mathrm{PBEsol}+U$ ($U$ = 3.9 eV).

\section{IV: SCF calculations on orthorhombic-\ch{LiMnO2}}

SCF calculations are performed on the experimentally refined structure of orthorhombic-\ch{LiMnO2} to determine the electronic density of states (DOS), magnetic moments, band gap, and electron density $\rho$, which are presented in Section 3B-C. An $8\times 8\times 8$ $\bm{k}$-point grid is used to sample the Brillouin zone. The functionals used are PBEsol, PBEsol+$U$, PBEsol+$U$+$V$, HSE06, and \rrscan+ $U$ ($U$ = 1.8 eV). The Hubbard $U$ and $V$ values within PBEsol are re-calculated self-consistently using DFPT at the experimentally refined configuration. The values are $U_{Mn}$ = 5.81 eV, $V$ = 0.76, 0.56, or 0.38 eV, and $U_{O}$ = 9.53 or 9.27 eV, which are in good agreement with the converged values listed in Table I in the main text. All SCF calculations are performed within QE, except for the \rrscan + $U$ calculations, which are done in VASP. The HSE06 SCF calculation is performed using optimized norm-conserving Vanderbilt (ONCV) PBEsol pseudopotentials selected from the PseudoDojo table \cite{pseudodojo-2018} (version 0.4.1) and a plane-wave basis set cutoff of 55 Ry (748 eV). ONCV pseudopotentials are chosen to minimize computational cost, as they are softer than PAWs.

\begin{table*}
\caption{Hubbard on-site U$_\mathrm{Mn}$, U$_\mathrm{O}$, and inter-site V (\ch{Mn}$3d$ - \ch{O}$2p$) parameters in orthorhombic \ch{LiMnO2}, self-consistently calculated using DFPT at the experimentally refined configuration and GGA+$U$+$V$ relaxed configuration (also shown in the main text Table I). Only the distinct values are listed. V of the short (1.9-2.0 \AA) and long (2.2-2.3 \AA) \ch{Mn}-\ch{O} bonds are shown in separate rows.}
    \centering
    \begin{adjustbox}{width=0.7\textwidth}
    \begin{tabular}{c | c | c}
    \hline 
         Hubbard & Orthorhombic & Orthorhombic \\
         Parameter [eV] & (exp) & (GGA+$U$+$V$ relaxed) \\
         \hline 
         $U_\mathrm{Mn}$ & 5.81 & 5.81 \\
         $U_\mathrm{O}$ & 9.27, 9.53 & 9.26, 9.55  \\
         $V$ (short) & 0.76, 0.56 & 0.72, 0.59 \\ 
         $V$ (long) & 0.38 & 0.38 \\
         \hline 
    \end{tabular}
    \end{adjustbox}
    \label{tab:hps}
\end{table*}

\section{V: Band structure calculations on orthorhombic-\ch{LiMnO2}}

The band structure of orthorhombic \ch{LiMnO2} (structure refined from ND) is calculated within $\mathrm{PBEsol}+U$, HSE06, and $G_0W_0$. The $\mathrm{PBEsol}+U$ band structure is computed using QE and its ground-state wave functions are used as a starting point for the $G_0W_0$ calculation. The $\mathrm{PBEsol}+U$ calculation is performed using ONCV PBEsol pseudopotentials selected from the PseudoDojo table (version 0.4.1)\cite{pseudodojo-2018} and a plane-wave basis set cutoff of 80 Ry (1088 eV). Norm-conserving pseudopotentials are chosen for these calculations as they are the only form of pseudopotential compatible with the BerkeleyGW package \cite{bgw-2012}. We re-calculate a self-consistent $U$ = 5.35 eV from DFPT, which is different from the value calculated using SSSP pseudopotentials. Variations in Hubbard parameters calculated using different pseudopotentials are expected \cite{timrov-hp-2022}. The HSE06 band structure is computed using VASP, due to the lower computational cost compared to QE. The reciprocal space discretization is chosen such that 3 $\bm{k}$-points separate each high symmetry point along the selected path through the Brillouin zone.

$G_0W_0$ calculations of the orthorhombic \ch{LiMnO2} band structure are performed using BerkeleyGW \cite{hybertsen-gw-1986, bgw-2012}. The static dielectric matrix ($\epsilon$) is computed within the random phase approximation using a $4\times 4 \times 4$ $\bm{q}$-point grid with a small shift of $0.001 \hat{z}$. The static $\epsilon$ is extended to finite frequencies using the Hybertsen-Louie plasmon pole model \cite{hybertsen-gw-1986}. A plane-wave cutoff of 80 Ry (1088 eV) and 1000 bands (60 occupied, 940 unoccupied) are used to generate $\mathrm{PBEsol}+U$ orbitals on a $4\times 4 \times 4$ $\bm{k}$-point grid.

\begin{figure}
    \centering
    \includegraphics[width=0.5\linewidth]{250323_pbesol_bands.pdf}
    \caption{Band structure of the experimentally refined ortho-\ch{LiMnO2} structure computed within PBEsol.}
    \label{fig:pbesol_bands}
\end{figure}

\begin{figure}
    \centering
    \includegraphics[scale=0.9]{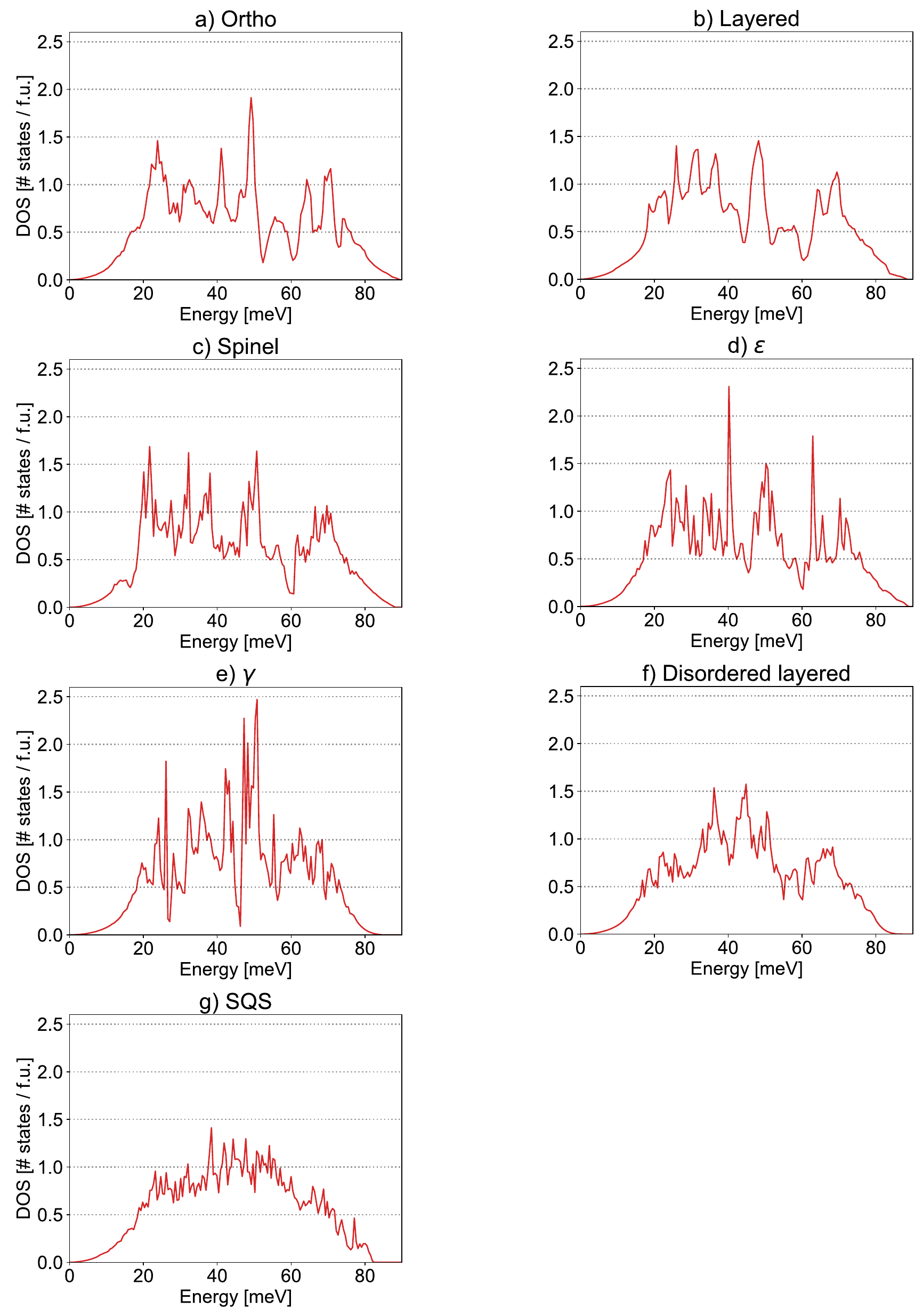}
    \caption{Phonon density of states of \ch{LiMnO2} phases.}
    \label{fig:enter-label}
\end{figure}

\begin{figure}
    \centering
    \includegraphics[width=0.65\linewidth]{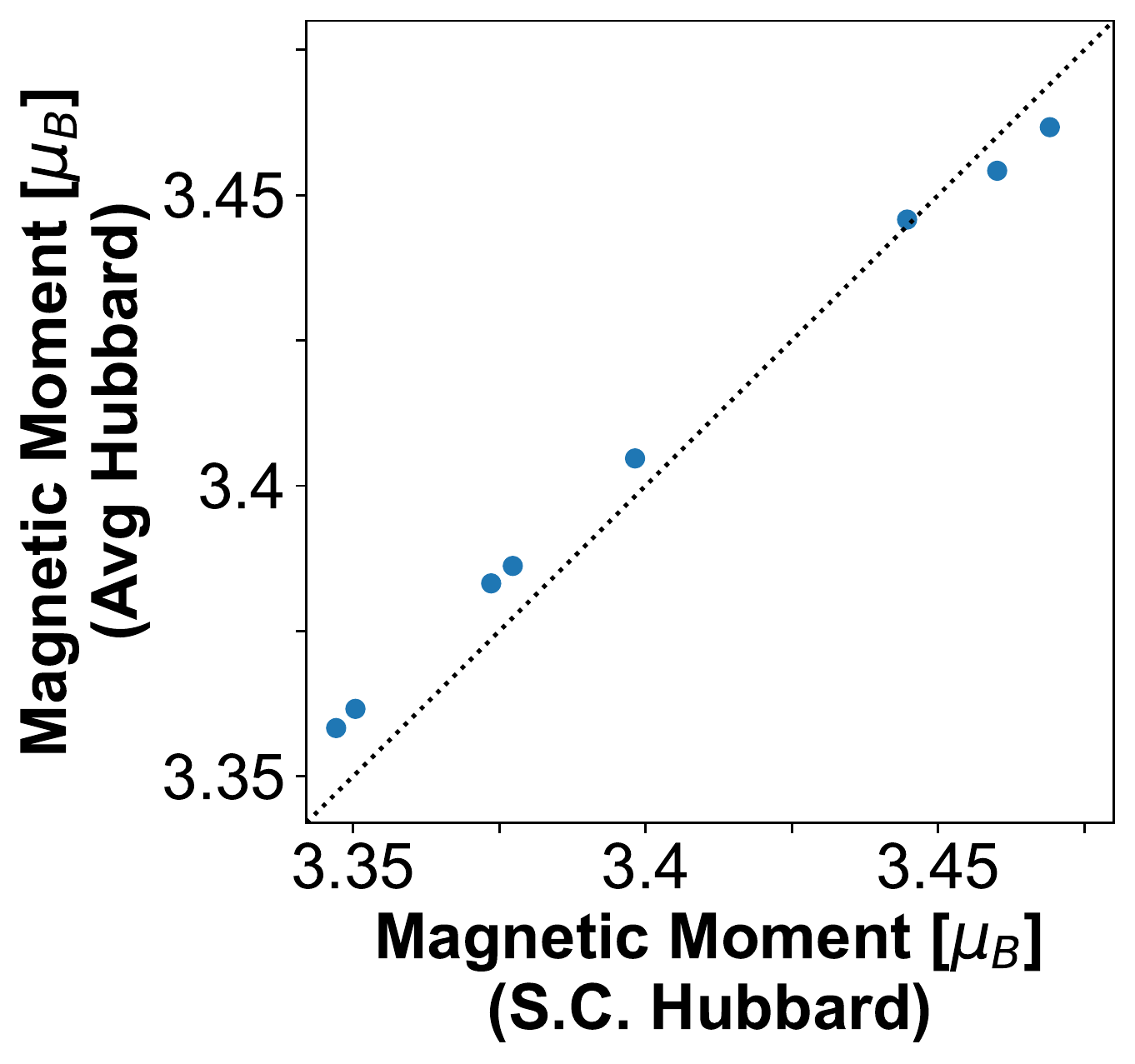}
    \caption{Comparing the Mn magnetic moments computed using PBEsol with self-consistent (sc) HPs and averaged (avg) HPs.}
    \label{fig:enter-label}
\end{figure}

\section{VI: Cluster expansion (CE) and Monte Carlo (MC) simulations}

\begin{figure}
    \centering
    \includegraphics[scale=0.65]{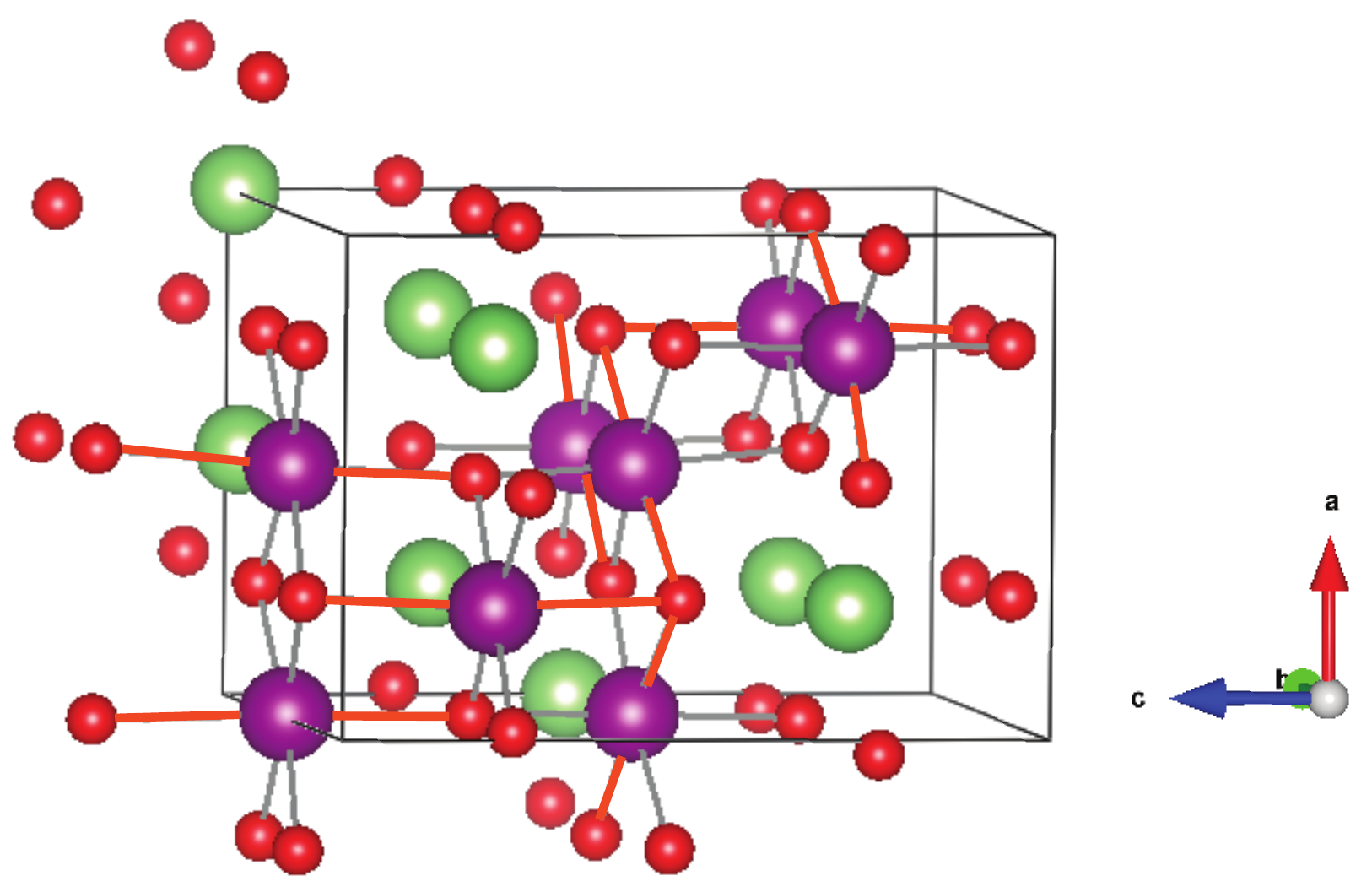}
    \caption{Special quasi-random structure of \ch{LiMnO2} relaxed with $\mathrm{PBEsol}+U$. JT axes are labeled in orange. Purple: Mn, green: Li, red: O.}
    \label{fig:sqs}
\end{figure}

We developed a CE to model the Li$^+$-Mn$^{3+}$-Mn$^{4+}$-O$^{2-}$ space on a rock-salt primitive cell of lattice parameter a = 2.97 $\AA$. The CE construction and MC simulations are performed using \texttt{smol} \cite{smol}. Pair, triplet, and quadruplet cluster interaction terms were included up to cutoffs of 9 $\AA$, 6 $\AA$, and 4.2 $\AA$, respectively. An electrostatic interaction term was included in the model to describe long-range ionic Coulomb interactions. The CE was trained on 182 rock-salt type structures in the \ch{LiMnO2}-\ch{Li2MnO3} pseudo-binary phase space. For each structure, we determine the AFM ground-state by enumerating 30 AFM orderings and relaxing each using PBEsol+$U$ ($U$ = 3.9 eV). From this structural geometry, we perform a static calculation using HSE06 to correct for the SIE and obtain more accurate \ch{LiMnO2} phase stability prediction. The $\epsilon$ phase was found by performing simulated heating of spinel \ch{LiMnO2} using canonical MC simulations. The special quasi-random structure (SQS) of \ch{LiMnO2} was stochastically generated using the CE and MC simulations \cite{zunger_sqs, stochastic_sqs_2013}. We relax the structure from the undistorted rock-salt lattice with lattice parameter a = 2.97 $\AA$, using PBEsol+$U$ ($U$ = 3.9 eV). The relaxed structure contains noncollinear JT distortions, as shown in Figure \ref{fig:sqs}.

\begin{figure}[!t]
    \centering
    \includegraphics[width=0.75\linewidth]{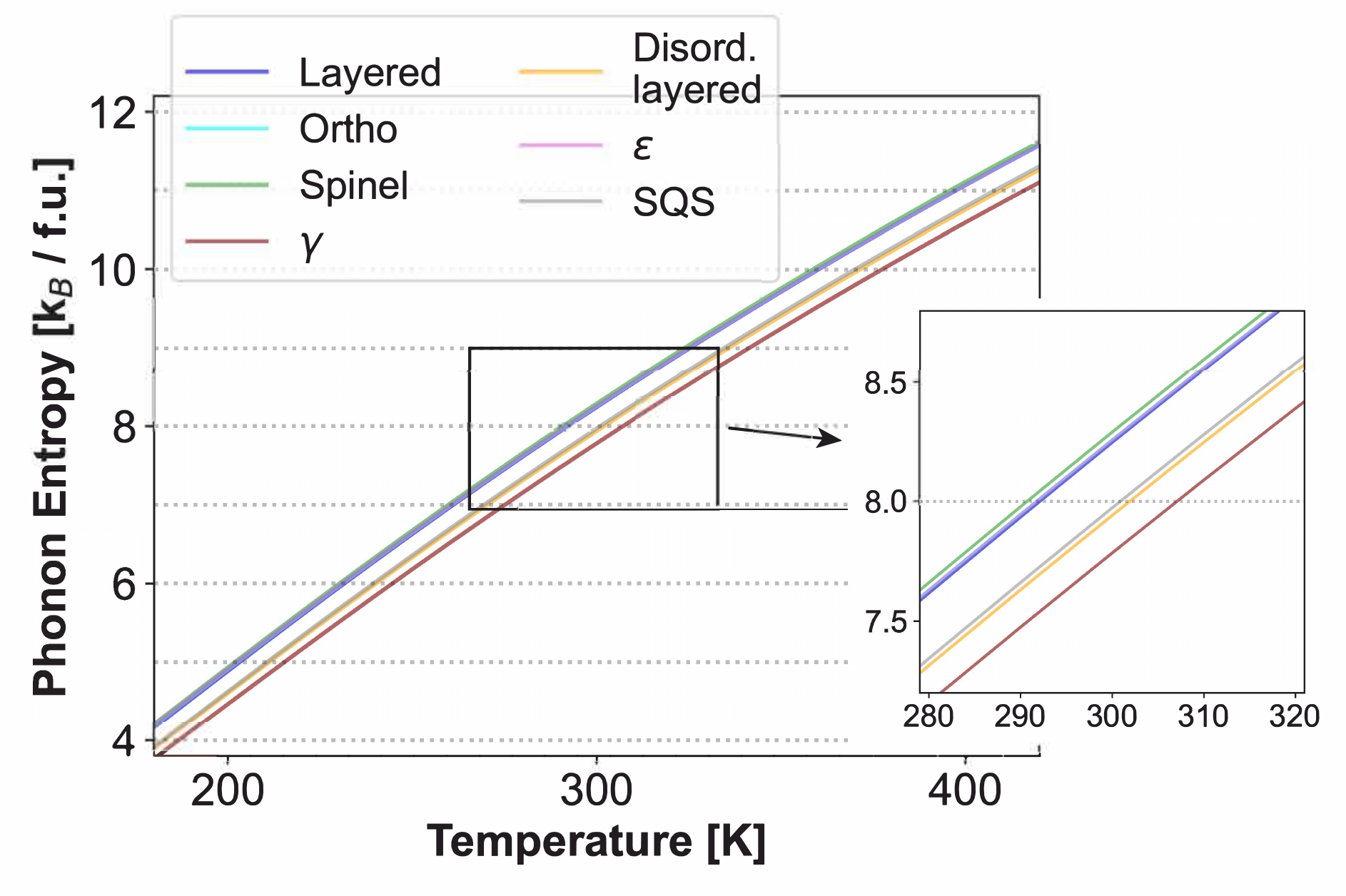}
    \caption{Phonon entropy as a function of temperature.}
    \label{fig:ph_entropy}
\end{figure}

\begin{figure}[!t]
    \centering
    \includegraphics[width=0.75\linewidth]{250518_lixmn2o4_hull_labeled.pdf}
    \caption{Formation energy of stoichiometric spinel (\ch{LiMn2O4}) relative to the fully delithiated spinel \ch{Mn2O4} and lithiated \ch{Li2Mn2O4} endpoints, computed using HSE06.}
    \label{fig:lixmn2o4}
\end{figure}

\begin{table}[]

    \centering
    \caption{Energy difference [meV/Mn] between the collinear and zig-zag ordering of JT axes in AFM layered \ch{LiMnO2} computed within several DFT functionals. A negative energy indicates that the collinear ordering is more stable.}
    \begin{adjustbox}{width=0.5\textwidth}
    \begin{tabular}{c|c|c|c}
    \hline
          PBEsol & \rrscan & PBEsol & HSE06 \\
          +$U$ & +$U$ & +($U$+$V$)$_{sc}$ & \\
          \hline
          -64 & -72 & -140 & -100 \\
         \hline
    \end{tabular}
    \end{adjustbox}
    \label{tab:layered_zz_ens}
\end{table}

\end{document}